\documentclass[pre,preprintnumbers,amsmath,amssymb,twocolumn]{revtex4}
\usepackage{graphicx}
\usepackage{bm,color}
\usepackage{tikz}

\newcommand{\nn}{\nonumber}

\newcommand{\rmd}{\mathrm{d}}
\newcommand{\rme}{e}

\renewcommand{\log}{\ln}

\begin{document}

\title{Perturbation Theory for Fractional Brownian Motion in Presence of Absorbing Boundaries}
\author{\bf  Kay J\"org Wiese$^{1}$, Satya N.\ Majumdar$^{2}$,  and Alberto
Rosso$^{2}$}

\affiliation{\medskip
$^{1}$CNRS-Laboratoire de Physique Th\'eorique de l'Ecole Normale Sup\'erieure, 24 rue Lhomond, 75005 Paris, France\\
$^{2}$CNRS-Laboratoire de Physique Th\'eorique et Mod\`eles Statistiques, 
Universit\'e Paris-Sud, 91405 Orsay, France.
\smallskip
}

\begin{abstract}
Fractional Brownian motion is a Gaussian process  $x(t)$  with zero
mean and two-time correlations $\langle x(t_1)x(t_2)\rangle =D \left (
t_1^{2H}+t_2^{2H}-|t_1-t_2|^{2H}\right)$, 
where $H$, with $0<H<1$ is called the Hurst exponent. For $H=1/2$, $x (t)$ is a Brownian motion, while for  $H\ne 1/2$, $x(t)$ is a non-Markovian process. Here we study $x (t)$
in presence of an absorbing boundary at the origin and focus on the probability density $P_+(x,t)$  for the process to arrive at $x$ at time $t$, starting near the origin at time $0$, given that it
has never crossed the origin. It has a scaling form $P_+(x,t)\sim
t^{-H}R_+(x/t^H)$. Our objective is to compute the scaling function
$R_+(y)$, which up to now was only known for the Markov case 
$H=1/2$. We develop a systematic perturbation
theory around this limit, setting $H=1/2+\epsilon$, to
calculate the  scaling function $R_+(y)$  to first order in $\epsilon$. We find that $R_+(y)$ behaves as $R_+(y)\sim y^{\phi}$ as $y\to 0$ (near the absorbing boundary), while
$R_+(y)\sim y^{\gamma} \exp ( -y^2/2)$ as $y\to \infty$, with  $\phi=1-4\epsilon+O(\epsilon^2)$ and $\gamma=1-2\epsilon+O(\epsilon^2)$. Our $\epsilon$-expansion
result confirms the scaling relation $\phi=(1-H)/H$ proposed in
Ref. \cite{phi1}. We verify our findings via numerical
simulations for $H=2/3$. The  tools  developed here are  versatile, powerful, and adaptable to different situations.  
\end{abstract}
\maketitle

\section{Introduction}
Survival of a species of bacteria, translocation of DNA 
through a nano-pore, and diffusion in presence of an absorbing
boundary are only few out of many situations, where the  central 
question is  the survival, or persistence of the underlying
stochastic process. More precisely, 
persistence, or  survival
probability $S(t)$ of a process is the probability that the process, starting from an initial positive position,
stays positive over a time interval $[0,t]$. For many stochastic processes
arising in non-equilibrium systems, persistence decays as a power law $S(t)\sim t^{-\theta}$, where $\theta$ is called the persistence 
exponent \cite{persistence}. 
For a simple
Markov process such as one-dimensional Brownian motion, $\theta=1/2$ \cite{redner}.
On the other hand, the exponent $\theta$ is non-trivial whenever the process is non-Markovian, i.e., 
has a memory.
In addition to  theoretical studies (for a brief review see
\cite{current}), the exponent $\theta$ has been measured in a number
of experiments
\cite{marcos,yurke,tam,walsworth,dougherty,leeuwen,stavans}. Even for
Gaussian non-Markovian processes,  $\theta$ is non-trivial
\cite{diffusionm}. For the latter processes that
are {\em close} to a Markov process (i.e., whose correlators are close to
that of a Gaussian Markov process) the exponent  $\theta$ was computed
perturbatively \cite{satyapert,braypert}.  This perturbation theory
has been used for various out-of-equilibrium systems, as  the global persistence at the critical point of the Ising model in $d=4-\epsilon$ dimensions \cite{global}, in  simple diffusion close to  dimension $0$ \cite{hilhorst}, and in fluctuating fields such as interfaces \cite{krug,cornell,brayinterface}.

\begin{figure}[t]\begin{center}%
{\includegraphics[width=8.6cm]{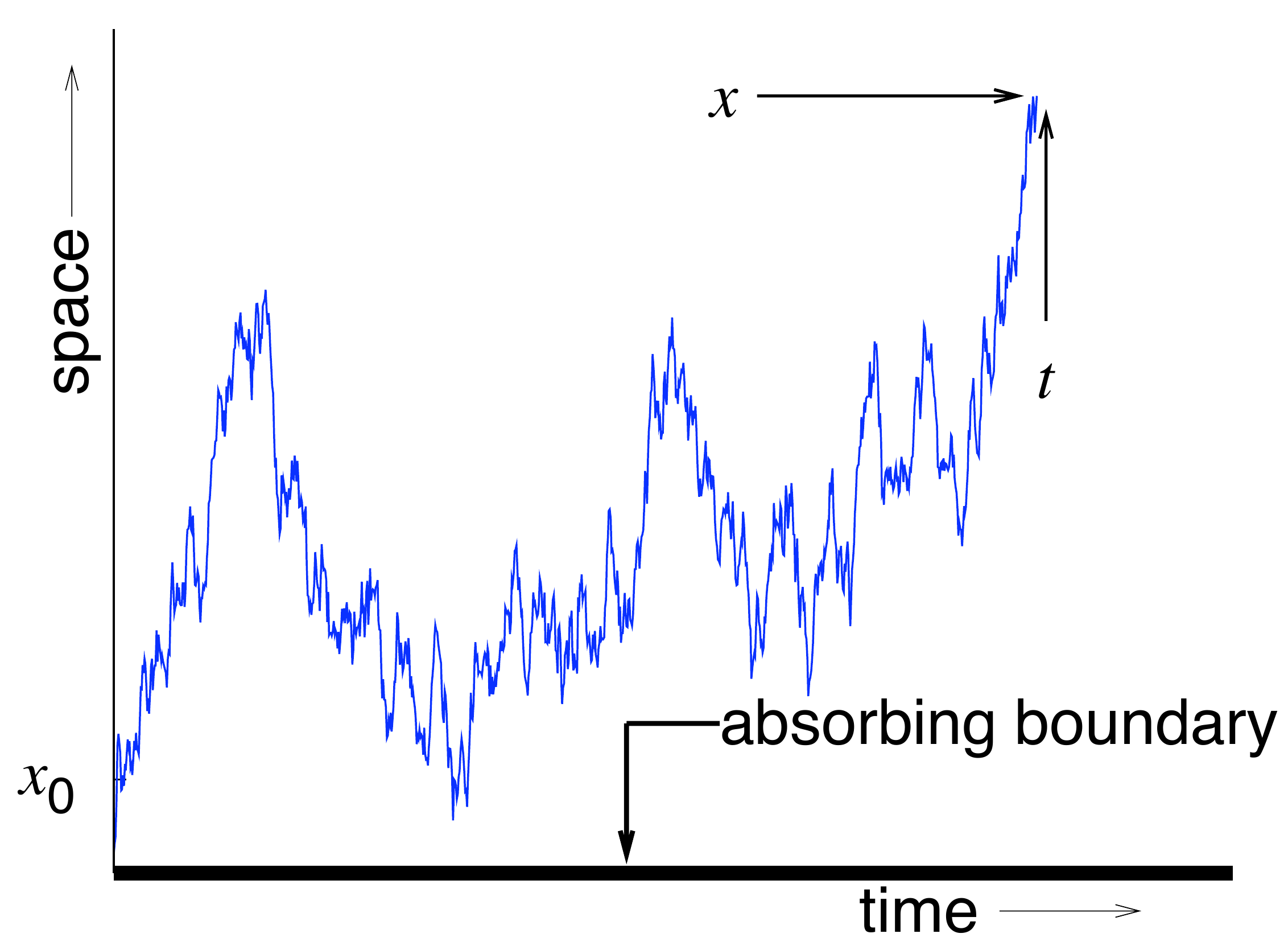}}%
\end{center}
\caption{(Color online) The fractional Brownian motion discussed in the main text.}
\label{a1}
\end{figure}%
A quantity that contains more spatial information than persistence $S(t)$ is the probability density $P_+(x,t)$
of the particle at position $x$ and at time $t$, given that it has survived (stayed positive) up to time $t$.  To investigate $P_+(x,t)$,
one can equivalently think  of a process 
on the positive semi-infinite line $[0,\infty]$ with  {\em absorbing}
boundary condition at the origin $x=0$ (see Fig.~\ref{a1}). 
The question is, how does $P_+(x,t)$ depend on $x$? In other words,
how does the presence of an absorbing boundary at the origin change the spatial dependence
of the probability density of the particle at time $t$? In particular, it is clear
that $P_+(x,t)$ must vanish as $x\to 0$ and $x\to \infty$. But how do
they vanish there? One of the main messages of our paper is that for
generic non-Markovian processes, 
$P_+(x,t)$ vanishes near its boundaries at $x=0$ and $x\to \infty$ in a non-trivial way, characterized
by non-trivial exponents. 

As the persistence $S(t)$, the probability $P_+(x,t)$ can be computed exactly for 
a Gaussian Markov process, as e.g.\  one-dimensional Brownian motion. For non-Markovian processes, even if they 
are Gaussian, $P_+(x,t)$ was not known. 
In this work, we consider $P_+(x,t)$ for a class of one-dimensional Gaussian processes known
as fractional Brownian motion (fBm), which are parametrized by their
Hurst exponent $H$, with  $0<H<1$.
The case $H=1/2$ corresponds to  ordinary Brownian motion, which is a Markov process, while
for $H\ne 1/2$ the process is non-Markovian. The purpose of this paper is to develop a systematic
perturbation theory to compute $P_+(x,t)$ for non-Markovian fBm's with
$H=1/2+\epsilon$, where $\epsilon$ is the expansion parameter for the
perturbation theory. Here we present the  result for $P_+(x,t)$
 to $O(\epsilon)$. It can be written as a combination of special
functions, i.e.\ error  and hypergeometric functions, see Eq.~(\ref{W}). 
To our knowledge, this is the first systematic
(exact up to $O(\epsilon^2)$) calculation of $P_+(x,t)$
for fractional Brownian motion with $H\ne 1/2$.%
\begin{figure}[t]
\begin{center}
{\includegraphics[width=8.6cm]{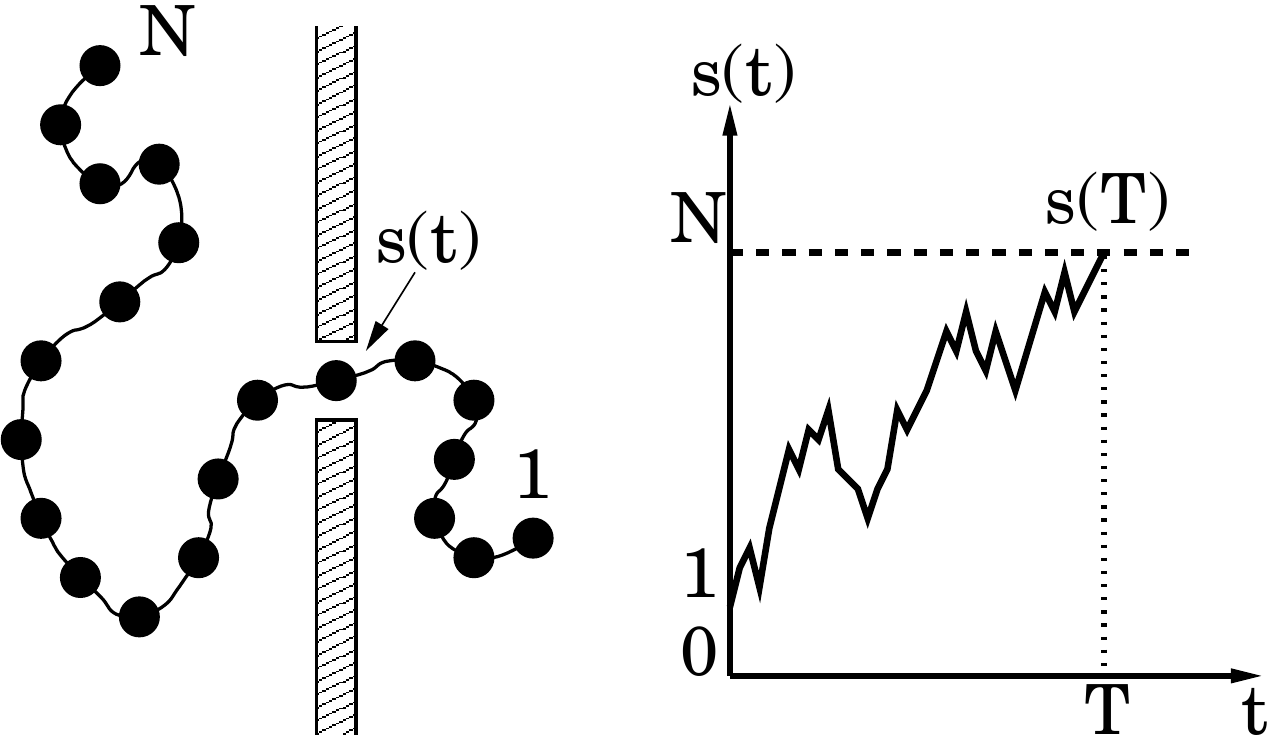}}
\end{center}
\caption{Left: Translocation of a polymer chain through a pore. Right: The translocation coordinate $s(t)$ denotes the number of the monomer that is crossing the pore at time $t$.}
\label{translocation}
\end{figure}%

Before detailing our results, let us position them into a broader context:
Fractional Brownian motion with $H\ne 1/2$ is relevant for polymer translocation through a nanopore. 
Consider a polymer chain composed of $N$ monomers  passing through a pore (translocation) from left to right, as drawn on Fig.~\ref{translocation}.
The dynamics of this translocation process has 
been investigated intensively due to its central role in understanding, e.g., viral injection 
of DNA into a host, or RNA transport through nano-pores, 
and mastering such applications as fast DNA or RNA sequencing through engineered channels \cite{ref1,ref2, ref3, ref4}. The 
translocation coordinate $s(t)$, namely the label of the 
monomer crossing the pore at time $t$, is key to quantitatively describing the translocation process \cite{ref5, ref6, ref7,ref7bis}, 
which begins when $s=1$, and ends when $s =N$, i.e., 
when the first and the last monomer of the chain enter 
the pore, respectively, see Fig.~\ref{translocation}.  For  large $N$, when the translocation is not yet  complete,  one can
view $s(t)$ as  a stochastic process on the semi-infinite line with
absorbing boundary conditions at $s=0$. The absorbing boundary at
$s=0$ models that if 
the chain  falls back to the left, i.e., on the starting side, it will
diffuse away and not try again. The quantity $P_+(s(t)=x,t)$ then represents the probability that $x$ monomers have translocated to the right at time $t$.  To model the process $s(t)$, one observes the following facts: (i) scaling arguments and numerical simulations show that $s(t)$ is subdiffusive \cite{kardar1}; (ii) in absence of boundaries, numerical simulations indicate that $s(t)$ is a Gaussian process \cite{kardar2}. Based on these observations it was proposed in Ref.~\cite{phi1} that a good candidate for $s(t)$ is a fractional Brownian motion with $H=1/(1+2 \nu)$, where the exponent $\nu$ describes the growth of the radius of gyration with the number of monomers ($R_g \sim N^\nu$) \cite{pgd}.  Thus for $\nu\ne 1/2$, $H<1/2$  and hence $s(t)$ is generically a non-Markovian process, with absorbing boundary conditions at $s=0$ and at $s=N$. Here we consider the limit of $N\to \infty$. Thus our results for $P_+(x,t)$ of a fBm with $H\ne 1/2$ are directly relevant for polymer translocation.

Directions for further applications are numerous: 
Recently a relation was established between the statistics of avalanches associated with
the motion of a driven particle in a
disordered potential  and persistence properties of the latter \cite{LeDoussalWiese2008a}. Higher-dimensional generalizations are avalanches of extended elastic objects, for which systematic field-theoretic treatments exist \cite{LeDoussalMiddletonWiese2008,LeDoussalWiese2008c,LeDoussalWiese2009a,LeDoussalWieseMoulinetRolley2009}. 
In few cases, no-hitting probabilities can be calculated for extended (non-directed) objects, as self-avoiding random walks avoiding extended objects \cite{LeDoussalWiese2008b}. Other approaches use real-space renormalization \cite{PLD1,PLD2}.

\medskip

This article is organized as follows: Since some of the computations are rather
technical, we first provide in Section \ref{summary} a brief summary of the main definitions and our principal results. In Section \ref {s:prelim},  we
introduce basic notations and reproduce  the known results for
$H=\frac12$. 
Section \ref{s:Perturbation theory} explains the basic
ideas of our perturbative approach, sketches the calculation, and
discusses some of the subtle points. Our predictions are compared to
numerical simulations in Section \ref{Numerics}. 
Conclusions are presented in Section \ref{s:Conclusions}.
More technical
points are relegated to two appendices: In Appendix \ref{a:action} the
correction to the action is derived.
 Appendix \ref{s:Evaluation of Z+} contains the explicit calculation
 of the perturbation theory. Finally, Appendix \ref{s:scaling} reviews  the arguments for
 the scaling law $\phi = (1-H)/H$.

\section{Summary of Definitions and Main Findings}
\label{summary}
Consider a particle, located at time $t=0$ at the origin $x=0$ and free to propagate on the real axis. For Gaussian processes, the probability  to find the particle inside the interval $(x, x+\rmd x)$ at time $t$ is given by
\begin{equation}
P(x,t)\, \rmd x= \frac{1}{\sqrt{2 \pi \langle x^2(t) \rangle}} e^{-\frac{x^2}{2 \langle x^2(t) \rangle} }\, \rmd x\ ,
\label{propagator1}
\end{equation}
where $\langle x^2(t) \rangle$ is the particle's mean square displacement.  A natural scaling variable is
\begin{equation}
y=\frac{x}{\sqrt{\langle x^2(t) \rangle}  }\ ,
\label{rescaled}
\end{equation}
and most of the  properties of the process are a function of this single variable. For example, the distribution probability in Eq.~(\ref{propagator1}) becomes
\begin{eqnarray}
P(x,t) \, \rmd x = R(y) \,\rmd y  \label{scaling}  \\
 R(y) =\frac{1}{\sqrt{2 \pi}} e^{- \frac{y^2}{2}}.
\label{propagator2}
\end{eqnarray}
In many problems the motion is confined to an interval, finite or
semi-infinite. In presence of {\em absorbing} boundaries,  the
probability distribution of the particle position, subject to the
condition that the particle has survived, has no longer a simple
Gaussian form since it has to vanish at the boundaries. However, one
can still express it as a function of the sole scaling variable $y$
defined in Eq.~(\ref{rescaled}),  where $\langle x^2(t) \rangle$ is
the particle's mean square displacement in the {\em unconstrained
(without boundaries)  process over the full real line}. In particular,
here we discuss the case where the particle can move on the positive
semi-axis and is absorbed whenever $x(t) < 0$.  We call $P_+(x,t)$ and
$R_+(y)$ with  $y$  given in Eq.~\eqref{rescaled}  the {\em normalized} probability distribution and the scaling function of the problem in presence of an absorbing boundary at the origin,
\begin{equation}
P_+(x,t)\, \rmd x= R_+(y) \, \rmd y\ .
\label{p+ry}
\end{equation}
  In contrast to the free case, the functional form of $R_+(y)$ is not the same for all Gaussian processes, but depends on the precise nature of the latter. Here we study a particular class of processes, the fractional Brownian motion (fBm), for which the autocorrelation function in absence of boundaries is
\begin{equation}\label{autocorr1}
\langle x(t_1) x(t_2)\rangle= D  \left( t_1^{2 H} +t_2^{2 H} -|t_1-t_2|^{2 H} \right)\ ,
\end{equation}
where $H$ with $0<H<1$ is the Hurst exponent. For $H=1/2$, the fBm identifies with Brownian motion
\begin{equation}
\langle x(t_1) x(t_2)\rangle= 2 D \min(t_1,t_2) \ ,
\label{autocorr2}
\end{equation}
where $D$ is the diffusion constant. Note that only for $H=1/2$, the Gaussian process $x(t)$ is Markovian. For other values of $H$, the process is 
non-Markovian.

For Brownian motion ($H=1/2$), the form of $R_+(y)$  can be obtained using the method of images (see Section \ref{s:prelim}), 
\begin{equation}
R_+^{(0)}(y)= y e^{-\frac{y^2}{2}}.
\label{BMR+}
\end{equation}
The superscript $(0)$ identifies the case $H=1/2$. For other values of $H$, due to the non-Markovian nature of the process,  the method of images no longer works and the computation of $R_+(y)$ becomes a challenging problem.  In this paper we  compute  this function,
using a perturbative approach for $H=1/2+\epsilon$, to first order in $\epsilon$. The final result is
\label{s:Numerics}\begin{figure}
\includegraphics[width=8.6cm]{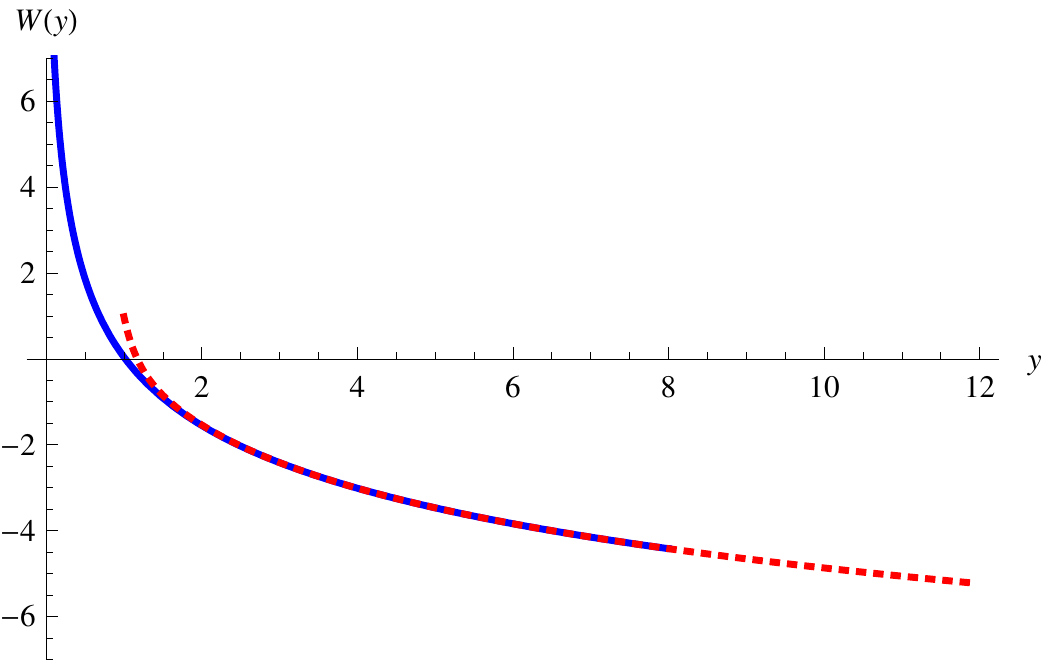}
\caption{(Color online) The function $W (y)$ defined in
Eq.~(\ref{W}). The  solid blue
line 
is the result of the series expansion (\ref{Wseries}), while the
dashed red line is the result of the asymptotic expansion
(\ref{Wasymp}). We  note that $W (1.01694)=0$. }
\label{f:W}
\end{figure}%
\begin{eqnarray}
\label{final1}
R_+(y)&=& R_+^{(0)}(y)  \left[1+\epsilon { W} (y)  +O(\epsilon^2) \right]   \\
\label{W}
W (y)&=&\frac{1}{6} y^4 \, _2F_2\!\left(1,1;\frac{5}{2},3;\frac{y^2}{2}\right)\nn \\
&& +\pi  (1-y^2)\,
   \text{erfi}\Big(\frac{y}{\sqrt{2}}\Big)+\sqrt{2 \pi } e^{\frac{y^2}{2}}
   y\nn \\
&& +\left(y^2-2\right) \left[\log 
   \!\big(2 y^2\big)+\gamma_{\mathrm{E}} \right]-3 y^2 \ ,
\end{eqnarray}
where $\gamma_\mathrm{E}$ is Euler's constant,
$_2F_2$ a hypergeometric
function, and $\text{erfi}$ the imaginary error-function.
We can write a {\em convergent} series-expansion
\begin{eqnarray}\label{Wseries}
W (y)&=& 4 y^4 \sum_{n=0}^{\infty} \frac{2^n n!\,
   y^{2n}}{(2 n+4)!} \nn \\
&& -\sum_{n=0}^{\infty}\frac{\sqrt{\pi } 2^{\frac{3}{2}-n} y^{2 n+1}}{(2
n-1) (2 n+1) n!}  \nn \\
&& +\left(y^2-2\right) \left[\log
   \left(2 y^2\right)+\gamma_{\mathrm{E}} \right]-3 y^2 
\end{eqnarray}
where each line is equivalent to the corresponding line in
Eq.~(\ref{W}). Both sums converge for all $y$, but problems of
numerical precision 
appear for $y>7$. In that region, one can use the  asymptotic
expansion 
\begin{equation}\label{Wasymp}
W (y)= 1-\gamma_{\mathrm{E}}- \log (2 y^{2}) +\frac{1}{2 y^2}-\frac{1}{2 y^4}+\frac{5}{4
   y^6}+O (y^{-8})
\end{equation}
At $y=7$, the difference between (\ref{Wseries}) and (\ref{Wasymp}) is
smaller than $10^{-6}$. 

We obtain, 
 at first order in
$\epsilon$,  the  asymptotic expansions of $R_{+} (y)$, 
\begin{eqnarray}
\label{final2}
R_+(y) & \xrightarrow{y \to 0}&  y \left[1  - 4 \epsilon  \log y - 2 \epsilon (\gamma_{\mathrm E}+\log 2  )   +\dots \right] \nonumber \\
R_+(y)& \xrightarrow{y \to \infty}&   y e^{-y^2/2} \left[ 1-2 \epsilon \log y +\epsilon(1-\log 2 -\gamma_{\mathrm E}) \right] \nn\\
&& +\dots 
\end{eqnarray}
These asymptotics can be recast into
\begin{align}
R_+(y) &\sim  y^\phi  &&\text{for} \quad y \to 0  \nonumber  \\
R_+(y) &\sim  y^\gamma e^{-\frac{y^2}{2}}  &&\text{for} \quad y \to \infty\ ,   
\label{fBmresult2}
\end{align}
where the two exponents
$\phi$ and $\gamma$ are at first order in
$\epsilon$ given by 
\begin{equation} 
 \label{final3}
\phi=1-4 \epsilon +O(\epsilon^2)\ , \quad
\gamma=1-2 \epsilon+O(\epsilon^2) \ .
\end{equation}
In a recent publication \cite{phi1} (reviewed  in Appendix
\ref{s:scaling}), a general scaling relation, valid for arbitrary
self-affine processes with stationary increments,  was proposed
between the exponent $\phi$, the persistence exponent $\theta$, and
the  Hurst exponent $H$,
\begin{equation}\label{scaling}
\phi=\frac{\theta} H\ .
\end{equation}
For fBm, it is known rigorously that $\theta=1-H$~\cite{krug}. This result  predicts that for fBm,
\begin{equation}\label{}
\phi=\frac{1-H}H\ .
\end{equation}
One of the  objectives of this paper was to verify this scaling relation up to $O(\epsilon)$ in a perturbation theory around $H=1/2$.
Using $H=1/2+\epsilon$, one expects 
$\phi=(1-H)/H = 1-4\epsilon +O(\epsilon^2)$ for fBm. This is in
agreement with our result (\ref{final3}),  
putting the scaling arguments on a firmer footing.

It is interesting to note that the scaling function $R_{+} (y)$ given
in Eq.~(\ref{propagator2}) has, at least  to $O(\epsilon)$, the same leading large-$y$ behavior
$\sim e^{-y^2/2}$ as in the unconstrained case
(\ref{propagator2}). 
 This   behavior can be  understood by a simple heuristic
 argument: far from the boundary the process is not ``aware'' of the
 latter. Our calculation reveals that the process nevertheless knows about the
 boundary, and   $R_+(y)$ 
 has a subleading power-law prefactor $y^{\gamma}$ where $\gamma$ is a
 new (independent) exponent, whose result to order $\epsilon$ is given in
 Eq.~(\ref{final3}).

Our analytical results are then verified via numerical simulations for
$H=2/3$.

\section{Preliminaries: Brownian case ($H=1/2$)}
\label{s:prelim}
To simplify notations, we set  $D=1$ in the following. The final result (\ref{final1}), expressed in the variable $y$, is of course independent of this choice.

The spreading of a Brownian particle is given by the Fokker-Planck equation 
\begin{eqnarray}
\partial _t  Z_+^{(0)}(x_0,x,t) &=& \partial_x^2 Z_+^{(0)}(x_0,x,t) \\
Z_+^{(0)} (x_0,x,t=0) &=& \delta (x-x_0)
\end{eqnarray}
The propagator $Z_+^{(0)}(x_0,x,t)$ times $ \rmd x$ gives the probability to find the Brownian particle inside the interval $(x, x+\rmd x)$ at time $t$, knowing that the particle was at $x_0$ at time $t=0$.  With absorbing boundary conditions at the origin we have, using the method of images
\begin{equation}\label{Z+0}
Z_+^{(0)}(x_0,x,t)=\frac{1}{\sqrt{4\pi t}}\left[e^{-(x-x_0)^2/4t} -e^{-(x+x_0)^2/4t}\right]\ . 
\end{equation}
This propagator is not a probability distribution because it is not
normalized. Its normalization, the so-called survival probability,
\begin{equation}
S(x_0,t) =\int_0^\infty \rmd x\, Z_+^{(0)}(x_0,x,t)= \text{erf}\left(\frac{x_0}{2 \sqrt{t}}\right)
\end{equation}
gives the probability that the particle is not yet absorbed by the boundary at $x=0$. The survival probability vanishes when $x_0\to 0$; however, in that limit, the probability distribution  for the  non-absorbed particles remains well-defined:
\begin{equation}\label{4}
 P_+^{(0)} (x,t)= \lim_{x_0 \to 0}\frac{Z_+^{(0)} (x_0,x,t)}{\int_0^\infty \rmd x \,Z_+^{(0)} (x_0,x,t)}\ .
\end{equation}
Another quantity with a finite limit for $x_0=0$ is
\begin{equation}\label{6}
 Z_+^{(0)} (x,t)= \lim_{x_0 \to 0}\frac{1}{x_0}  Z_+^{(0)} (x_0,x,t) =  \frac{x e^{-\frac{x^2}{4 t}}}{2 \sqrt{\pi } t^{3/2}}\ .
\end{equation}
This allows to write the  probability  $ P_+^{(0)} (x,t)$ as
\begin{equation}\label{f10}
 P_+^{(0)} (x,t)=\frac{Z_+^{(0)}(x,t)}{ \int_{0}^{\infty}\rmd x\,  Z_+^{(0)}(x,t) }=\frac{x}{2t} e^{-\frac{x^2}{4 t}}.
\end{equation}
Using in  Eq.~(\ref{f10}) the scaling variable defined in (\ref{rescaled}), $y=x/\sqrt{2 t}$, we recover (\ref{BMR+}).  Eq.~(\ref{f10}) is simpler than Eq.~(\ref{4}) because the $x_0$ dependence is discarded from the beginning. We will use this definition to compute  $Z_+ (x,t)$ for $H=1/2+\epsilon$.

\section{Perturbation theory ($H\neq 1/2$)}
\label{s:Perturbation theory}
The process $x(t)$ is Gaussian for all values of $H$, but it is
Markovian only for $H=1/2$. For all other values of $H$, the
process is non-Markovian and this makes the problem difficult to
solve. Our idea is to expand around $H=1/2$. In a first step, we
construct an action, which calculates expectation values of the
Gaussian process $x(t)$, with bulk expectation values
(\ref{autocorr1}). In a second step, we obtain the propagator with
absorbing boundary conditions at $x=0$. In a third step we 
calculate the probability $P_+(x,t)$ perturbatively, using the action
constructed in step 1. In the fourth step, we put together all pieces
and interpret our
result. 

 \subsection{Step $1$: The Action} 
 For all $H$, $x(t)$ is a Gaussian process, therefore  the statistical weight of a path $x(t')$ without any boundary is proportional to $\exp(-{\cal S}[x])$ where the action ${\cal S}[x]$ is quadratic in $x$ and given by
\begin{equation}
\label{action0}
{\cal S}[x]=\int_0^t  d t_1 \int_0^t  d t_2 \, \frac12 x(t_1) G(t_1,t_2) x(t_2). 
\end{equation}
Note that we use standard field-theoretic notation, noting $f(x)$ a
function of the variable $x$, and ${\cal S}[x]$ a functional,
depending on the function $x(t')$, with $0<t'<t$. 

The kernel $G(t_1,t_2)$ of the action is related to the auto-correlation function of the process via
\begin{equation}
\label{GA}
 G^{-1}(t_1,t_2)=\langle x(t_1) x(t_2) \rangle.    
\end{equation}
For $H=1/2$, the action is simple. In this case, setting $D=1$,
\begin{equation}
\label{GAB}
 [G^{(0)}]^{-1}(t_1,t_2)=\langle x(t_1) x(t_2) \rangle = 2 \min(t_1, t_2). 
\end{equation}
Using the result (\ref{G0}) in Eq.~(\ref{action0}), we recover the standard Brownian action 
\begin{equation}
{\cal S}^{(0)}[x]= \frac14 \int_0^t d t' \left(   \partial_{t'} x \right)^2.
\label{S0}
\end{equation}
For a generic value of $H$ the kernel $G(t_1,t_2)$ becomes non-local. For $H=1/2+\epsilon$ one can write
\begin{equation}
\label{expans1}
{\cal S}[x]={\cal S}^{(0)}[x]+\epsilon\, {\cal S}^{(1)}[x] +\dots
\end{equation}
where ${\cal S}^{(0)}[x]$ is the action (\ref{S0}) and ${\cal S}^{(1)}[x]$ has been computed in 
 Appendix \ref{a:action}
\begin{eqnarray}
\label{S1}
{\cal S}^{(1)}[x]&=& - \frac12 \int_0^t  d t_1 \int_{t_{1}}^t  d t_2 \, \frac{\partial_{t_1} x(t_1) \partial_{t_2} x(t_2)}{|t_1-t_2|} \nn\\&& - 2 {\cal S}^{(0)}[x] (1+\log \tau)\ .
\end{eqnarray}
Note that we have introduced a regularization for coinciding times $t_1=t_2 \rightarrow \log|t_1-t_2|=\log \tau$  where $\tau>0$ is
the UV cutoff. A first-principle definition would necessitate a
discretization in time. It is  however sufficient to check that the law (\ref{autocorr1}) is correctly reproduced, and that the final result is cutoff independent.

 \subsection{Step $2$: The Propagator with an Absorbing Boundary}
 For a generic value of $H$, the propagator $Z_+(x_0, x, t)$,  denoting the probability that the particle reaches $x$ at time $t$, starting from $x_0$ at time $0$, and staying positive over the interval $[0,t]$,
 can be written using standard path integral notation as 
\begin{equation} 
 Z_+(x_0, x ,t)= \int_{x(0)=x_0}^{x(t)=x} {\cal{D}}[x] \,\rme^{-{\cal S}[x]}  \, \Theta[x]\ .
\end{equation}
Here $\Theta[x]$ is an indicator function that is $1$ if the path $x(t')$ stays positive over the interval $[0, t]$ and $0$ otherwise. The action ${\cal S}[x]$ is given in (\ref{action0}). 
 In the limit $x_0\to 0$, we expect, as in the Brownian case ($H=1/2$), the propagator to vanish
as $x_0^{\phi_0}$ where the  yet unknown exponent $\phi_0$ depends on $H$. Note that for $H=1/2$, $\phi_0=1$ (see Eq.~(\ref{6})). For $H=1/2+\epsilon$, we expect that
$\phi_0=1+a_1 \epsilon+ O(\epsilon^2)$, where $a_1$ is yet
unknown.
 Analogous to Eq.~(\ref{6}) for $H=1/2$ we define $Z_+(x,t)$ as
\begin{equation}\label{32}
 Z_+(x,t)= \lim_{x_0 \to 0}\frac{1}{x_0^{\phi_0}} \int_{x(0)=x_0}^{x(t)=x}  {\cal{D}}[x] \,\rme^{-{\cal S}[x]}\,  \Theta[x]\ .
\end{equation}
Using the expansion of the action given in Eq.~(\ref{expans1})  and $\phi_0=1+a_1 \epsilon$, we write to leading order in $\epsilon$
\begin{eqnarray}
\label{Z+def}
 \lefteqn{Z_+(x,t)}\nn \\
&=&\lim_{x_0 \to 0}\frac{1}{x_0^{1+a_1 \epsilon}} 
 \int_{x(0)=x_0}^{x(t)=x} {\cal{D}}[x] \left(1-\epsilon {\cal S}^{(1)}[x]\right) \rme^{-{\cal S}^{(0)}[x]}   \, \Theta[ x]     \nonumber \\
 &=& \lim_{x_0 \to 0} \Big\{ Z_+^{(0)}(x,t)\left[1 -a_1 \epsilon \log(x_0) \right] \nn \\
&& \ \qquad - \frac \epsilon {x_0} \int_{x(0)=x_0}^{x(t)=x} {\cal{D}}[x]\, {\cal S}^{(1)}[x] \, \rme^{-{\cal S}^{(0)}[x]}   \, \Theta[ x] \Big\}  \nonumber \\ &=& Z_+^{(0)}(x,t) + \epsilon Z_+^{(1)}(x,t)\ ,
\end{eqnarray}
 where $ Z_+^{(0)}(x,t)$ is defined in Eq.~(\ref{6}) and
 $Z_+^{(1)}(x,t)$ is 
\begin{eqnarray}
\label{Z1def}
Z_+^{(1)}(x,t)=   \lim_{x_0 \to 0} &\!\Big\{\! & \frac{-1}{x_0} \int_{x(0)=x_0}^{x(t)=x} {\cal{D}}[x] \, {\cal S}^{(1)}[x] \, \rme^{-{\cal S}^{(0)}[x]}   \,\Theta[ x]  \nn\\&&  -a_1  \log(x_0)  Z_+^{(0)}(x,t) \Big\}
\end{eqnarray}
We will see that for $\phi_0=1- 4 \epsilon$, i. e. $a_1=-4$,  $Z_+^{(1)}(x,t)$ is independent of $x_0$.

\subsection{Step $3$: Calculation of $Z_{+}^{(1)} (x,t)$}\label{step3}

The main achievement of this paper is the calculation of
$Z_+^{(1)}(x,t)$ defined in Eq.~(\ref{Z1def}). This calculation is rather involved, both
conceptually and technically. Therefore, we will relegate several technical  calculations to Appendix
\ref{s:Evaluation of Z+}. 
Eq.~(\ref{Z1def}) can be devided into three pieces:
\begin{eqnarray}
\label{eqZ1}
 Z_+^{(1)}(x,t)
& =&Z_+^A(x,t)  \\&&  + \lim_{x_0 \to 0}\Big[ Z_+^B(x_0,x,t)  -a_1  \log(x_0)  Z_+^{(0)}(x,t)   \Big] \nn \\
 \label{eqZA} 
 Z_+^A(x,t)&=&2(1+\log \tau)  \\
&& \times \lim_{x_0 \to 0}\frac{1}{x_0}\int_{x(0)=x_0}^{x(t)=x} {\cal{D}}[x] {\cal S}^{(0)}[x]\, \rme^{-{\cal S}^{(0)}[x]}   \,\Theta[ x]   \nn \\ \label{eqZB}
 Z_+^B(x_0,x,t)&=&   \frac{1}{4}\int_0^t\int_0^t \rmd t_1 \rmd t_2\\
&& \times \frac{1}{x_0} \int_{x(0)=x_0}^{x(t)=x} {\cal{D}}[x]
 \frac{\dot{x(t_1)} \dot{x(t_2)}  }{|t_1-t_2|}\, \rme^{-{\cal S}^{(0)}[x]}   \,\Theta[ x]  \nn
\end{eqnarray}
The first term, $Z_+^A(x,t)$ is simple, and is evaluated in appendix
\ref{s:Evaluation of Z+A}.  
We now come to the evaluation
of the contribution $Z_{+}^{B} (x_0,x,t)$, defined in Eq.~(\ref{eqZB}). 
In Fig.~\ref{f:Z+} we show a path which contributes to $Z_+^B(x_0,x,t)$. The sum of all these paths is a product of
transition probabilities. Explicitly, it reads, ordering
$t_{1}<t_{2}$, which gives an extra factor of 2 compared to (\ref{eqZB}):
\begin{eqnarray}\label{weight}
\lefteqn{\!\!\!Z_+^B(x_0,x,t)} \nn\\
&=& \frac{1}{2 x_0}    \int\limits_{0}^{t} \rmd t_2
\int\limits_{0}^{t_2}\rmd t_1  \int\limits_{x_{1}>0}\, \int\limits_{\tilde
x_{1}>0}\, \int\limits_{x_{2}>0}\,  \int\limits_{\tilde x_{2}>0}
\nn \\
&& \Big[
Z_+^{(0)}(x_0,\tilde x_1,t_1) D({\tilde x}_1,x_{1}) \frac{Z_+^{(0)}( x_1,x_{2}, t_2-t_1)}{|t_2-t_1|} \nn \\
&& \times  D(x_2,{\tilde x}_2)Z_+^{(0)}(\tilde x_2,x,t-t_2) \Big]\ .
\end{eqnarray}
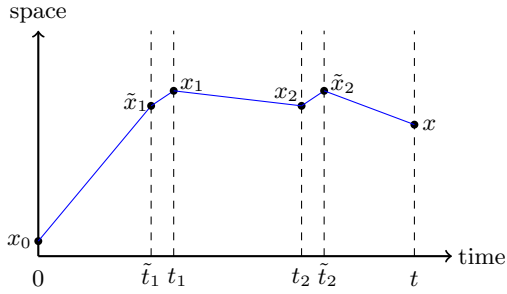
\begin{figure}
{{\begin{tikzpicture}
\draw [->,thick] (0,0) -- (5.5,0);
\draw [->,thick] (0,0) -- (0,3);
\draw [dashed] (1.5,0) -- (1.5,3);
\draw [dashed] (1.8,0) -- (1.8,3);
\draw [dashed] (3.5,0) -- (3.5,3);
\draw [dashed] (3.8,0) -- (3.8,3);
\draw [dashed] (5,0) -- (5,3);
\fill (0,.2) circle (1.5pt);
\node (x) at (-.25,.2) {$x_{0}$};
\node (x) at (0,-.3) {$0$};
\draw [blue] (0,.2) -- (1.5,2.);
\fill (1.5,2.) circle (1.5pt);
\node (tx1) at (1.5-.2,2.05) {$\tilde x_{1}$};
\node (tt1) at (1.5,-.25) {$\tilde  t_{1}$};
\draw [blue] (1.8,2.2) -- (1.5,2.);
\fill (1.8,2.2) circle (1.5pt);
\node (tx) at (1.8+.25,2.28) {$x_{1}$};
\node (t1) at (1.85,-.28) {$ t_{1}$};
\fill (3.5,2.) circle (1.5pt);
\node (x2) at (3.5-.2,2.15) {$ x_{2}$};
\node (t2) at (3.5,-.28) {$t_{2}$};
\draw [blue] (3.8,2.2) -- (3.5,2.);
\fill (3.8,2.2) circle (1.5pt);
\node (tx2) at (3.8+.25,2.3) {$\tilde x_{2}$};
\node (tt2) at (3.85,-.25) {$ \tilde t_{2}$};
\draw [blue] (1.8,2.2) -- (3.5,2.);
\fill (5,1.75) circle (1.5pt);
\node (tf) at (5,-.3) {$t$};
\node (delta) at (5.2,1.75) {$x$};
\draw [blue] (5,1.75) -- (3.8,2.2);
\node (t) at (5.9,0) {time};
\node (space) at (0,3.2) {space};
\end{tikzpicture}}}\caption{Graphical representation of the
path-integral for $Z_{+}^{B} (x_0,x,t)$ given in Eq.~(\ref{Z1def}).}\label{f:Z+}
\end{figure}%
$Z_+^{(0)}(\tilde x_1,x_0,t_1)$, $Z_+^{(0)}(x_1,
x_2, t_2-t_1)$  and $Z_+^{(0)}(\tilde x_2,x,t-t_2)$  are
defined in (\ref{Z+0}). The factors $D(x_1,\tilde x_1)$  and
$D(x_2,\tilde x_2)$ take into account the terms $\partial_{t
_1}x(t_1)$ and $\partial_{t _2} x(t_2)$ in the action $S_1[x]$.
\begin{eqnarray}\label{Q}
D(\tilde x_1, x_1) &=& \lim_{ \rmd t \to 0} \frac{\left(
x_1-\tilde x_1\right)}{\rmd t} Z_+^{(0)}( x_1,\tilde x_1,\rmd t) \nn \\
& = &   \lim_{ \rmd t \to 0} \frac{\left(  x_1-\tilde x_1\right)}{\rmd t} \frac{e^{-\frac{(\tilde x_1-x_1)^2}{4 \rmd t}}}{\sqrt{2 \pi \rmd t}}\nn \\
&=&  -2  \delta'( x_1- \tilde x_1).
\end{eqnarray}
Finally, we have set $\tilde t_{1}=t_{1}$, and $\tilde
t_{2}=t_{2}$, since we have taken the limit of their differences to
0. 
In order to perform the six integrations in Eq.~(\ref{weight})
it turns out to be convenient to  evaluate its Laplace transform,
$\tilde{Z}_+^B(x_0,x,s)$. 
From now on, we will always denote with  $\tilde f (s)$ the 
Laplace transform of a function $f (t)$, 
defined as 
\begin{equation}
\tilde{f} (s):=\int_0^\infty \rmd t \,e^{-s t}\,f (t)\ .
\label{LT}
\end{equation}
This Laplace transform leads to two important simplifications: 
The first simplification is that now the nested time-integrals over
$t_{1}$ and $t_{2}$ become a product. To see this, we remind that if
$f_{1}$ and $f_{2}$ are two functions which depend on $t$, then the
Laplace transform of their convolution is simply the product of their
Laplace transforms,
\begin{eqnarray}\label{a6}
&&\!\!\!\int_{0}^{\infty}\rmd t\, \rme^{-t s}\left[ \int_{0}^{t} \rmd t_{1}
f_{1} (t_{1}) f_{2} (t-t_{1})  \right] \nn \\
&&= \int_{0}^{\infty}\rmd t\, \int_{0}^{\infty} \rmd t_{1}\, \int_{0}^{\infty }\rmd t_{2}\,  \delta (t-t_{1}-t_{2}) \nn \\
&&\qquad \times 
f_{1} (t_{1})\rme^{-t_{1} s} f_{2} (t_{2})\rme^{-t_{2}s}
\nn \\
&&= \left[ \int_{0}^{\infty} \rmd t_{1} f_{1} (t_{1})\rme^{-t_{1} s}
\right]\left[ \int_{0}^{\infty} \rmd t_{2} f_2 (t_{2})\rme^{-t_{2} s}
\right] \nn \\
&&= \tilde f_{1} (s) \tilde f_{2} (s)\ . \label{conv-theorem}
\end{eqnarray}
This consideration generalizes to 3 and more times.

We obtain for the Laplace transform of (\ref{weight})
\begin{eqnarray}\label{ZBLP1}
\tilde{Z}_+^B(x_0,x,s)&=&-  \frac{2}{x_0} \int\limits_{x_{1}>0}\int\limits_{x_{2}>0}
\tilde{Z}_+^{(0)}(x_0,x_1,s)   \tilde{Z}_+^{(0)}(x_2,x,s)
\nn \\
&&\times  \partial_{ x_1}\partial_{ x_2}\left[  \int_0^\infty  \rmd
t\, e^{-s t}  \frac{Z_+^{(0)}(x_1,  x_2,t)}{t} \right]  \qquad      
\end{eqnarray}
The second simplification is even more important, and is most easily
understood on the example of the bulk propagator
\begin{equation}\label{B11}
Z^{(0)} (x,y,t) := \frac{e^{-(x-y)^2/4t}}{\sqrt{4\pi t}} \ .
\end{equation}
Its Laplace-transform is 
\begin{equation}\label{B12}
\tilde Z^{(0)} (x,y,s)= \frac{1}{2 \sqrt{s}}e^{-\sqrt{s} |x-y|}\ .
\end{equation}
While integrals over  $x>0$ involving (\ref{B11}) give
error-functions, which are hard to integrate further, the same
integrals over (\ref{B12}) remain similar exponential functions; the only
complication is that one has to distinguish between
$x$ smaller or larger than $y$.

To evaluate (\ref{ZBLP1}), we now have to calculate the
Laplace-transforms of its factors: 
\begin{eqnarray}\label{46}
\tilde{Z}_+^{(0)}(x,y,s)&=&\frac{e^{-\sqrt{s}  | x-y |}-e^{-\sqrt{s}  (x+y)}}{2 \sqrt{s} } \ .
\end{eqnarray}
Finally, the term in brackets in Eq.~(\ref{ZBLP1}) can be rewritten, using a
Fourier decomposition for $Z_+^{(0)}(x_2,  x_1,t)$, as 
\begin{eqnarray}
\lefteqn{ \int_0^\infty  \rmd t\, e^{-s t}  \frac{Z_+^{(0)}(x_1,  x_2,t)}{t} } \nn \\
&=& \int_{-\infty}^{\infty} \frac{\rmd k}{2 \pi}  \int_0^\infty  \rmd t\, \frac{e^{-(s+k^2) t}}{t}\left[ e^{i k (x_1-x_2)}-e^{i k (x_1+x_2)}\right] \nonumber \\
 &=& -\int\limits_{-\infty}^{\infty} \frac{\rmd k}{2 \pi}  \left[ e^{i k
 (x_1-x_2)}-e^{i k (x_1+x_2)}\right]\left[ \log([s{+}k^2]\tau) +\gamma_{\mathrm E} \right] \nn \\
\label{48}
\end{eqnarray}
Note that the time integral  in the second line of Eq.~(\ref{48})  is
  diverging at small times.   Since the path integral is defined as discretized in
  time, a natural approach consist in discretizing  this integral, with a
step-size $\tau$. This would  indeed be the only possible approach for
stronger divergences, like $1/t^{2}$. However, since our integral is
only logarithmically diverging, we can take an easier path,  by using a small-time cutoff $\tau$:
\begin{eqnarray}\label{a7}
\lefteqn{\int_0^\infty \frac{e^{- (s +k^2) t}}{t} \rmd t}  \\
&& \to \int_\tau^\infty  \frac{e^{- (s +k^2) t}}t \rmd t= - \log([s +k^2]
\tau) -\gamma_{\mathrm E} +O(\tau) \ .\nn  
\end{eqnarray}
We note that the regularization by discretization gives the same
result apart from the term  $-\gamma_{\mathrm{E}}$.  We will check later
 that it only contributes to the normalization, which will drop
from the final result.

Collecting the results of Eqs.\ (\ref{46}) and (\ref{48}) in
Eq.\ (\ref{ZBLP1}), and doing the remaining space-derivatives we find
\begin{eqnarray}
\tilde{Z}_+^B(x_0,x,s) &=&     \frac{2}{x_0} \int\limits_{-\infty}^{\infty}   \frac{\rmd k}{2 \pi}  k^2 \left[ \log (\tau (s+k^2)) +\gamma_{\mathrm E} \right] \nn \\
& &\times  \int\limits_{x_{1}>0}\int\limits_{x_{2}>0}    \left[ e^{i k (x_1-x_2)}+e^{i k (x_1+x_2)}\right]  \nonumber \\
&&  \times \frac{e^{-\sqrt{s}  | x-x_2 |}-e^{-\sqrt{s}  (x_2+x)}}{2 \sqrt{s} }   \nn\\
&& \times  \frac{e^{-\sqrt{s}  | x_0-x_1 |}-e^{-\sqrt{s}  (x_1+x_0)}}{2 \sqrt{s} }
\label{ZBLP2}
\end{eqnarray}
Performing the space-integrations, we find 
\begin{eqnarray}\label{a8}
\lefteqn{\tilde{Z}_+^B(x_0,x,s)}  \nonumber \\
&=&    \frac{4}{x_0} \sqrt{s} \int\limits_{-\infty}^{\infty}  \frac{\rmd k}{ 2 \pi}
\left[ \cos( k x_0)- e^{-\sqrt{s} x_0}\right] \left[ \cos( k x)- e^{-\sqrt{s} x}\right] 
\nn \\
 &&\qquad\qquad \times \frac{  k^2 [\log(\tau(s+k^2))+\gamma_{\mathrm E}] }{(s+k^2)^2} \qquad
\end{eqnarray}
Note that this is (rescaling $k \to \sqrt{s} k$)
\begin{eqnarray}\label{km11}
{\tilde{Z}_+^B(x_0,x,s)} & =&  \frac4{x_0} \int\limits_{-\infty}^{\infty}\frac{\rmd
k}{2\pi} \Big[\!\cos( k x\sqrt{s}  )- 
e^{-x \sqrt{s} }\Big] \nn\\
&&\qquad \quad \times \Big[\!\cos( k x_0 \sqrt{s}  )- 
e^{-x_0 \sqrt{s} }\Big] \nn\\
&& \qquad \quad \times\frac{ k^2 \left[ \log \big(\tau s (1+k^{2} )
\big)+\gamma_{\mathrm E} \right]}{\sqrt{s} (1 + k^2)^2 }    \qquad 
\end{eqnarray}
The next step is to invert this Laplace transform which is performed
in Appendix \ref{AB2}. 

\subsection{Step $4$: The
Probability $P_+(x,t)$ }
The final result for $Z_{+}^{(1)} (x,t)$ is given in Eqs.~(\ref{Z1})
and (\ref{A+B1+B2}) of Appendix \ref{sub:Sum}, expressed in terms of the scaling variable $z=x/\sqrt{2 t}$. Note that setting $\phi_0=1-4 \epsilon$, i. e. $a_1=-4$, the term  $Z_{+}^{(1)} (x,t)$ does not depend on 
$x_0$:
\begin{eqnarray}
\frac{ Z_+^{(1)}(z,t) }{ Z_+^{(0)}(z,t)}&=&  ( z^2-2) \left[  \log (2 z^{2 }t) +\gamma_{\mathrm E}
\right]-2  +{\cal I} (z) + c (t) \nn \\
c (t)&= & \log (t)-2 \gamma_{\mathrm{E}} +2\   
 \label{Z1}
\end {eqnarray}
 where ${\cal{I}}(z)$ is defined in Eq.~(\ref{Iz}). The first line is
 arranged as to not contribute to the normalization, whereas  $c(t)$  is independent of $z$ and will not appear in the final conditional probability.  $\gamma_{\mathrm E}$ is Euler's constant. 
The probability distribution, $P_+(x,t)$, to find a  non-yet-absorbed particle in the interval $(x, x+ \rmd x)$ can be computed following the lines of 
 Eq.~(\ref{f10}) to order $\epsilon$ as
 \begin{align}
&P_+(x,t)= \frac{Z_+^{(0)}(x,t) +\epsilon Z_+^{(1)}(x,t)}{\int\limits_0^\infty \! \rmd x\, \left( Z_+^{(0)}(x,t) +\epsilon Z_+^{(1)}(x,t))\right)   } \nonumber \\
&=\frac{Z_+^{(0)}(x,t) }{\int\limits_0^\infty \! \rmd x\, Z_+^{(0)}(x,t)}\!\left[ 1+\epsilon \left( \frac{ Z_+^{(1)}(x,t)}{ Z_+^{(0)}(x,t)}  - \frac{\int\limits_0^\infty \! \rmd x\, Z_+^{(1)}(x,t)}{\int\limits_0^\infty \! \rmd x\, Z_+^{(0)}(x,t)}  \right)\!\right]
\label{cp0}
\end{align}
Note that   the term proportional to  $c (t)$  cancels in
normalized objects such as $P_+(x,t)$. Therefore, we obtain
\begin{align}\label{cp1}
&P_+(x,t) \, \rmd x = R_+^{(0)}(z) \rmd z\\
&\qquad\quad \times \Big\{ 1+\epsilon \Big[(z^2-2)\left( \gamma_{\mathrm E} +\log(2 z^2 t)\right)-2 +{\cal{I}}(z)  \Big] \Big\} \, \nn
\end{align}
where $R_+^{(0)}(z)=z \exp(-z^2/2)$, and ${\cal{I}}(z)$ is given in Eq.~(\ref{Iz}). The result in Eq.~(\ref{cp1}) still involves both $z$ and $t$. The reason is that  for $H\neq 1/2$  the natural scaling variable  is $y=x/(\sqrt{2} t^{1/2+\epsilon})$ instead of $z=x/\sqrt{2 t}$, as can be seen  from Eq.~(\ref{rescaled}). 
To rewrite  Eq.~(\ref{cp1}) in terms of $y= zt^{\epsilon}$, we note
that 
\begin{eqnarray}\label{}
R_{+}^{(0)} (z)\rmd z &=& R_{+}^{(0)} (y t^{\epsilon}) t^{\epsilon} \rmd y\nn \\
&=&  R_{+}^{(0)} (y) \left\{1+ \epsilon \left[\frac{y \partial_{y}
R_{+}^{(0)} (y)}{  R_{+}^{(0)} (y)} +1 \right] \ln t\right\} \nn \\
&=&  R_{+}^{(0)} (y)\left\{1- \epsilon \left[y^{2}-2 \right]\ln t \right\}\ .
\end{eqnarray}
This gives for Eq.\ (\ref{cp1}) up to terms of order $\epsilon^2$
\begin{align}\label{cp1y}
&P_+(x,t) \, \rmd x = R_+^{(0)}(y) \rmd y\\
&\qquad\quad \times \Big\{ 1+\epsilon \Big[(y^2-2)\left( \gamma_{\mathrm E} +\log(2 y^2 )\right)-2 +{\cal{I}}(y)  \Big] \Big\} \, \nn
\end{align}
This is the final result announced in equation (\ref{final1}), with
${\cal I}(y)$ calculeted in (\ref{Iz}) and below.

\section{Comparison to numerics} \label{Numerics}
In this Section, we compare our analytical results with numerical
simulations. More specifically, we consider the super-diffusive process  
with $H=\frac{2}{3}$.

\begin{figure}[t]\begin{center}%
{\includegraphics[width=8.6cm]{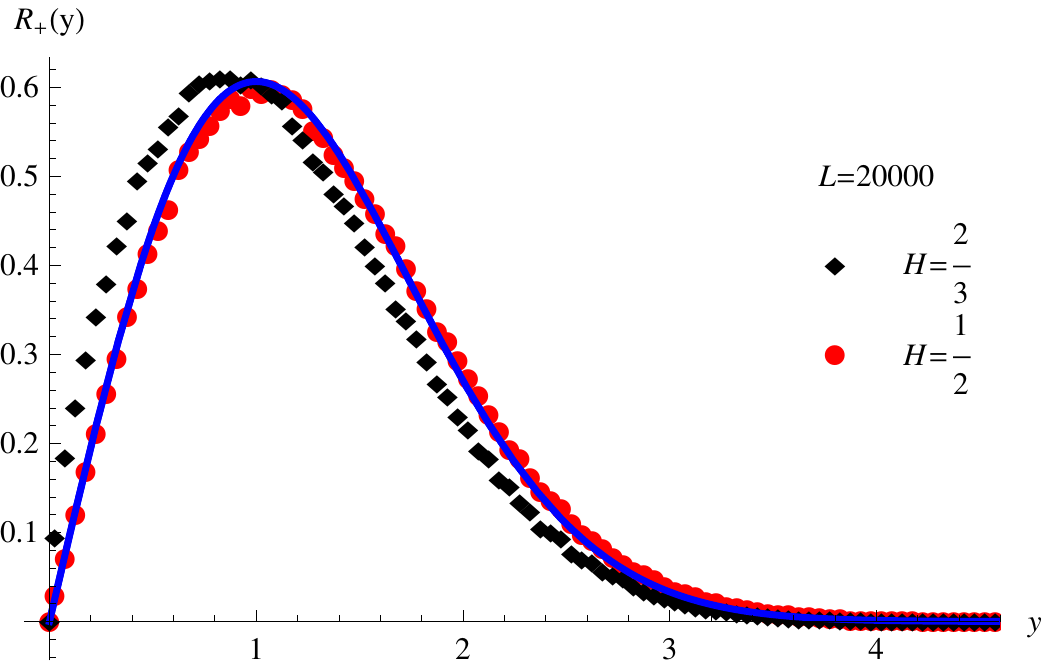}}%
\end{center}
\caption{(Color online) $R_{+} (y)$. Analytical result for  Brownian
motion $R_{+}^{(0)} (y) = y \rme^{-y^{2}/2}$(solid
blue line), and simulation data for $L=20000$, and  $H=\frac{1}{2}$ (red dots), as
well as for the fBm with $H=\frac{2}{3}$ (black diamonds). Histograms are performed over $4 \times 10^5$ paths.  }
\label{histlin}
\end{figure}%

\subsection{Methodology of simulations}
We aim to sample a fBm processes $x(t)$ at discrete times $t_1=1,t_2=2,\ldots,t_L=L$. 
The covariance matrix  of  $\{  x_1, \ldots , x_i, \ldots ,x_L\}$ coincides with the autocorrelation function of the original fBm process in Eq.~(\ref{autocorr1}), setting $D=1$, 
\begin{equation}
 C_{i,j}=\langle x_i x_j \rangle=  i^{2 H}+j^{2H}-|i-j|^{2 H} \ .
\label{num1}
\end{equation}
The $L \times L$ covariance matrix $C$ is symmetric and has positive eigenvalues; it is thus possible to find a matrix $A$,  positive and symmetric, such that $C=A^2$. Matrix $A$ is called the square root of $C$.

One can simulate paths of a fBm using the standard procedure for Gaussian  correlated processes: (i) Determine $A$, the square root of $C$. (ii) Each path $\vec x=\{ x_1, \ldots , x_i, \ldots ,x_L\}$ is given by the matrix multiplication  $\vec x=A \vec\eta$. The vector  $\vec \eta =\{\eta_1,\eta_2,\ldots,\eta_L\}$ is a set of $L$ independent Gaussian numbers with unitary variance and zero mean. It is easy to check that these paths are characterized by  the correct covariance matrix (\ref{num1}).

Unfortunately this procedure is time consuming, as for step (i)  it requires  the full diagonalization of C. Better results are obtained by making use of the stationarity of the increments $ \xi_i = x_{i}-x_{i-1}$ (we set $x_1=\xi_1$). Using Eq.~(\ref{num1}) we can compute ${\tilde{C}}$,  the covariance matrix of the increments,
\begin{equation}
{\tilde{C}}_{i,i+ k}:= \langle\xi_i \xi_{i+k} \rangle=
|k-1|^{2 H}+(k+1)^{2 H}-2\, k^{2 H}  \ ,
\end{equation}
where $k=0,\ldots,L-i$, and ${\tilde{C}}_{i+k,i}={\tilde{C}}_{i,i+
k}$. The matrix ${\tilde{C}}$ is  symmetric and positive definite like
the matrix $C$, but it also is a Toeplitz matrix. For Toeplitz
matrices efficient numerical methods allow to avoid the full
diagonalization of ${\tilde{C}}$.  In particular, the Levinson
algorithm (for a practical implementation of Levinson's algorithm see
\cite{simulation} and \cite{Garcia}) is suitable for  first passage problems, as  it recursively generates the increment $\xi_{i+1}$ given
$\xi_1,...,\xi_i$.  The points of the fBm path are given by $x_i=\sum_{j=1}^{i} \xi_j$. In our simulation we are interested only in positive paths ($ x_i>0$ for all $ i$). The Levinson method allows to discard negative paths whenever a $x_i<0$ is generated, without building the full path.

\begin{figure}[t]\begin{center}%
{\includegraphics[width=8.6cm]{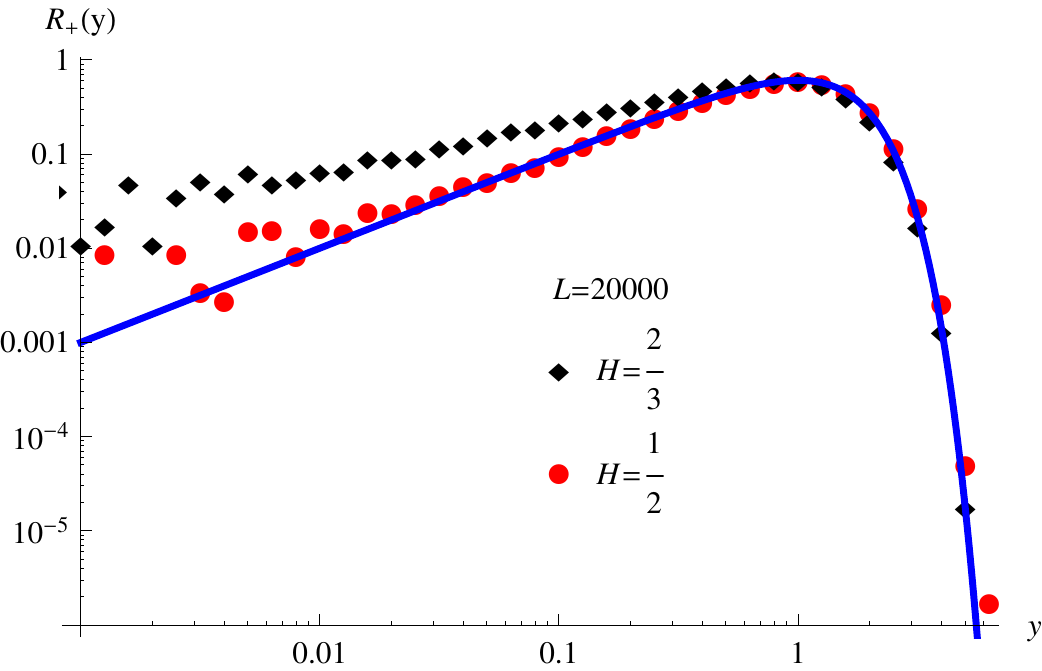}}%
\end{center}
\caption{(Color online) $R_{+} (y)$. Analytical result for  Brownian
motion $R_{+}^{(0)} (y) = y \rme^{-y^{2}/2}$(solid
blue line), and simulation data for $L=20000$, and  $H=\frac{1}{2}$ (red dots), as
well as for the fBm with $H=\frac{2}{3}$ (black diamonds).  Histograms are performed over $4\times 10^5$ paths. }
\label{histlog}
\end{figure}%

\subsection{Simulation results}

\begin{figure}[t]
\centerline{\includegraphics[width=8.6cm]{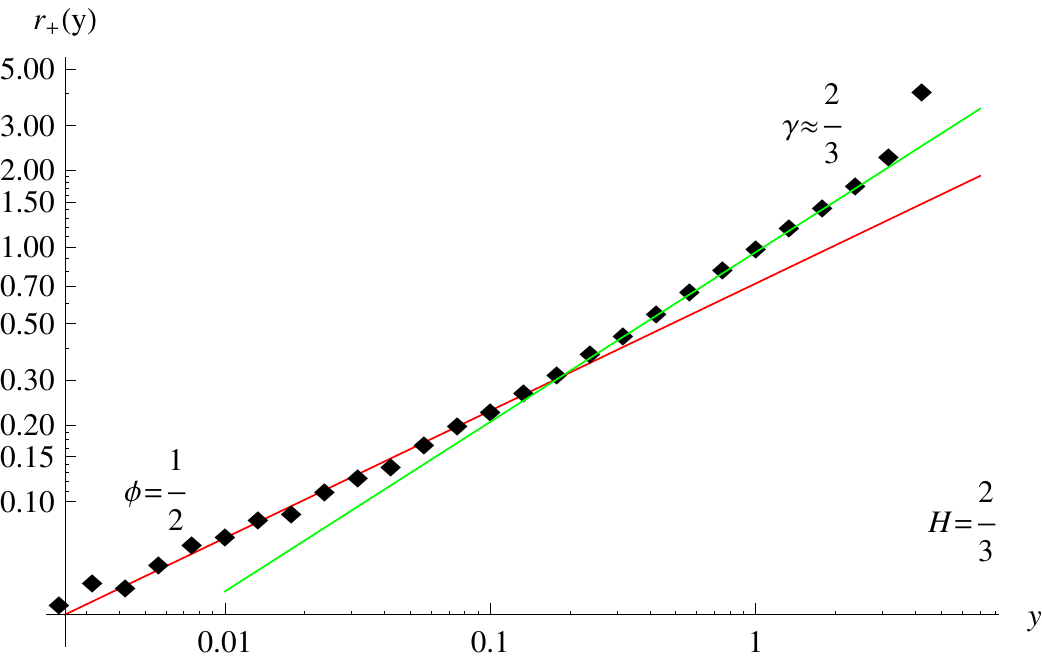}}
\caption{(Color online) The numerically determined function $r_{+} (y)$, defined
in Eq.\ (\ref{r+def}) for the fBm with $H=\frac{2}{3}$ (black
diamonds), using $L=20000$ and  $4\times 10^5$ paths. The asymptotic small-$y$ behavior is
consistent with $\phi=\frac{1-H}H=\frac12$.   The large-$y$
asymptotics (including the amplitude) was taken from (\ref{r-theory}),
with slope $\gamma\approx 1-2\epsilon=\frac23$. }
\label{f:H=2/3}
\end{figure}

For each positive path we record the final position $x_L$. The histogram of the rescaled variable $y:=x_L/(2 L^H)$ is the scaling function $R_{+} (y)$. The results for $H=2/3$ and the Markovian case $H=1/2$ are presented on Figs.\ \ref{histlin} and \ref{histlog}. For small $y$ the scaling function, $R_{+} (y)$ behaves as a power-law, with
an exponent $\phi$. For $H=1/2$ we expect $\phi=1$, for $H=2/3$ we expect $\phi=1/2$.  Inspired by our perturbative calculation we predict that for $y\to \infty$, $R_{+} (y)$ behaves like $\sim y ^\gamma e^{-y^2/2}$. 
In order to facilitate the comparison, we define the scaling function
\begin{equation}\label{r+def}
r_{+} (y):=\rme^{\frac{y^{2}}{2}} R_{+} (y)\ .
\end{equation}
The numerical data for the scaling
funtion $r_{+} (y)$ defined in Eq.~(\ref{r+def}) are shown on
Fig.\ \ref{f:H=2/3} for $H=2/3$. They clearly show two distinctive power-law
behaviors: For small $y$ this power law is $\sim y^{\phi}$ with $\phi =\frac{1}{2}$,
predicted by the scaling relation $\phi =\frac{1-H}{H}$. For large
$y$ a larger exponent $\gamma = 0.7 \pm 0.03$ is measured. This is consistent with the perturbative calculation, which suggests $\gamma>\phi$ for $H>1/2$ and $\gamma<\phi$ for $H<1/2$.

\begin{figure}[t]
\centerline{\includegraphics[width=8.6cm]{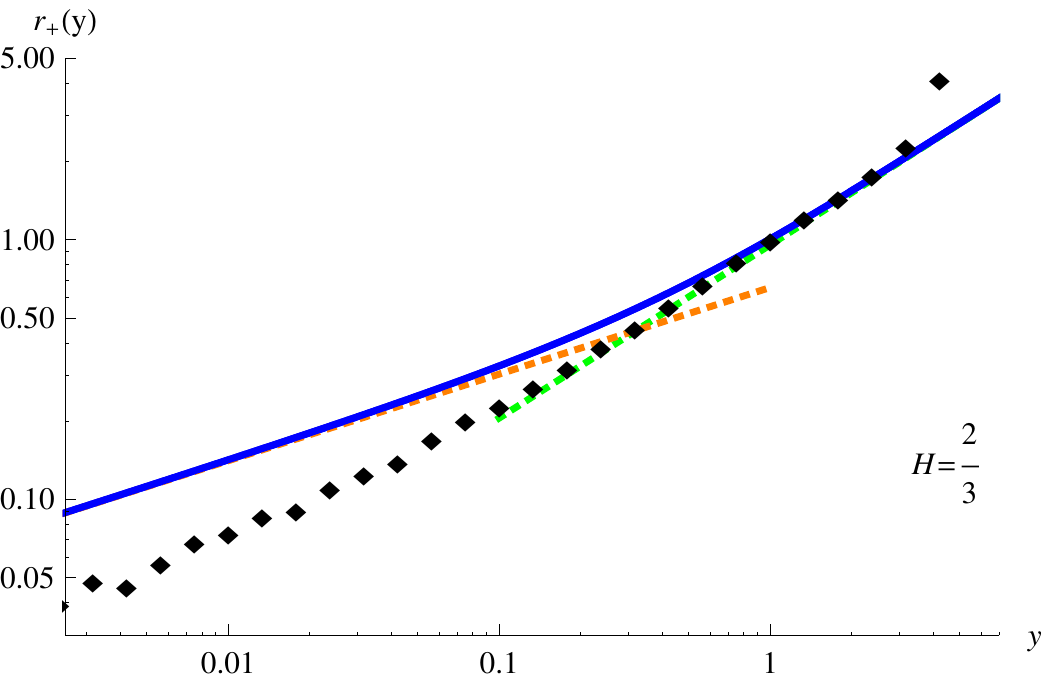}}
\caption{(Color online) Blue solid line: The function $r_{+}^{\epsilon} (y)$, defined
in Eq.\ (\ref{r-theory}) for $H=\frac{2}{3}$, i.e.\ $\epsilon
=\frac{1}{6}$. The dashed lines are the predicted asymptotic
behaviors, $\sim y^{\phi}$ (for small $y$) and $\sim y^{\gamma}$ (for large $y$). Superimposed are the simulation data  shown on Fig.\ \ref{f:H=2/3}. Note that there is no fitting parameter. The deviation at small $y$ is due to the fact that $\epsilon $ is rather large, so the order-$\epsilon$ slope $\phi\approx 1-4\epsilon =\frac13 $ is smaller than the exact result $\phi=\frac{1-H}{H}\equiv\frac{ 1/2-\epsilon}{1/2+\epsilon}$, which evaluates to $\frac 12$. For large $y$, but smaller than $3$, the effective cutoff of the simulation, both amplitude and slope are correctly predicted.}
\label{f:H=2/3-theory}
\end{figure}

A more accurate comparison between the numerical data and the perturbation theory is possible. Our perturbative result  given in Eq.\ (\ref{final1}) is equivalent to $r_{+}(y) = y \left[1+\epsilon W (y) +O (\epsilon^{2}) \right] $. In order to compare to numerics, we use 
\begin{equation}\label{r-theory}
r_{+}^{\epsilon} (y) = y\, \rme^{\epsilon W (y)} +O (\epsilon^{2})\ .
\end{equation}
While the two expressions are  equivalent to order $\epsilon$, the
latter 
(\ref{r-theory}) has the merit to resum the logarithms for small and
large $y$ into the power-law behavior 
\begin{equation}\label{44}
r_{+}^{\epsilon}(y) \sim  \left\{\begin{array}{cl}
y^{\phi_\epsilon} & \qquad \mbox{for }y \to 0\\
y^{\gamma_\epsilon} & \qquad\mbox{for }  y \to  \infty 
\end{array} \right.\ ,
\end{equation}
where the exponents are the order-$\epsilon$ results
\begin{equation}
 \label{final3+}
\phi_\epsilon=1-4 \epsilon \ , \quad
\gamma_\epsilon=1-2 \epsilon  \ .
\end{equation}
For $H=2/3$, i.e.\ $\epsilon=1/6$, we predict a scaling  $\sim y^{\gamma_\epsilon}$, $\gamma_\epsilon =
\frac{2}{3}$, using (\ref{final3+}).  Note that the curve drawn is exatly the asymptotic
behavior of our analytical result (\ref{r-theory}), using (\ref{W}),
thus also the {\em amplitude} and not only the exponent are estimated. This can more clearly be seen on Fig.~\ref{f:H=2/3-theory}, where the solid (blue) line represents the
theoretical order-$\epsilon$ prediction, and the dashed line the
asymptotic behaviors given in Eq.~(\ref{44}). 

\begin{figure}[t]
\centerline{\includegraphics[width=8.6cm]{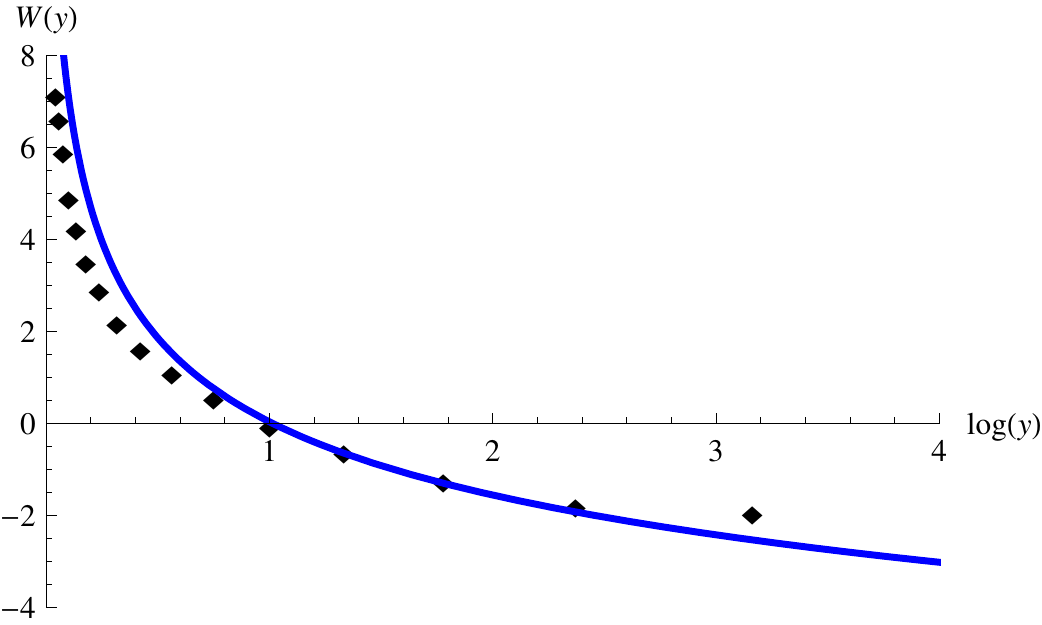}}
\caption{(Color online) Blue solid line: The function $W (y)$, defined
in Eq.\ (\ref{W}). Black diamonds: Estimation of $ W
(y)$ from the  numerical data for $H=\frac{2}{3}$, using relation (\ref{W-extract}).  The
agreement is quite good for $1\le y\le 3$. It
breaks down for  $y>3$ due to numerial
problems. For $y<1$, the deviations can be attributed to the large
value of $\epsilon$.}
\label{f:W2}
\end{figure}

Conversely,  relation (\ref{r-theory}) can be used to extract $W
(y)$ from $r_{+} (y)$, 
\begin{equation}\label{W-extract}
W (y) \approx \frac{1}{\epsilon} \log\! \left(\frac{r_{+} (y)}{y} \right)\ .
\end{equation}
This relation should work the better, the smaller $\epsilon$ is. Using our numerical results for $H=\frac23$, we obtain the curve presented on Fig.~\ref{f:W2}.   The
agreement is quite good for $1\le y\le 2.5$. It
breaks down for larger $y$ due to numerial
problems. For $y<1$, the deviations can be attributed to the large
value of $\epsilon$.


\section{Conclusions}\label{s:Conclusions}
In this article, we develop a systematic scheme to calculate
the corrections to the universal scaling function $R_{+} (y)$ for fractional Brownian motion, in an $\epsilon= H-\frac{1}{2}$ expansion.
 We  compute the full scaling function $R_{+} (y)$  to  first order in $\epsilon$.  
In particular we find that $R_+(y)$ behaves as $R_+(y)\sim y^{\phi}$ as $y\to 0$ (near the absorbing boundary), while
$R_+(y)\sim y^{\gamma} \exp ( -y^2/2)$ as $y\to \infty$ (far from the boundary), with, at the first order in $\epsilon$, $\phi=1-4\epsilon+O(\epsilon^2)$ and $\gamma=1-2\epsilon+O(\epsilon^2)$. 
For small $\epsilon$ our results confirm the scaling relation found in Ref.~\cite{phi1}: $R_+(y)\sim y^{\phi}$ with  $\phi=\theta/H$. For fractional Brownian motion it is known that $\theta=1-H$, so that $\phi=(1-H)/H\approx 1-4 \epsilon+\ldots$.
Far from the boundary, i.e.\ for large $y$,  the leading behavior
$R_+(y)\sim \exp ( -y^2/2) $  recovers the Gaussian propagator
\eqref{propagator2} in absence of boundaries; our  approach shows that
$R_+(y)$ has a subleading power law prefactor $y^\gamma$, where
$\gamma$ is a new (independent) exponent.

Our numerical simulations   show that the predictions of the
asymptotic behavior of $R_+(y)$  hold at $H=2/3$. In particular the two exponents $\gamma$ and $\phi$ have been measured and shown in Fig.~\ref{f:H=2/3}.

Let us stress that  few results are  known about  non-Markovian
processes in presence of boundaries. Perturbation theory developed in
this paper can provide substantial new insight here.
The method is  versatile and can in principle be extended to the
calculation of other quantities  such as the propagator for a process
confined to a finite interval with absorbing boundaries, or
alternatively with other, e.g.\ reflecting boundary conditions. Particularly
interesting for applications would be the hitting
probability $Q(x,L)$, the probability that a generic stochastic
process starting at $x$ and evolving in a box $[0, L]$ hits the upper
boundary at $L$ before hitting the lower boundary at $0$ \cite{hitting}. In the context of polymer translocation, the hitting probability is the probability that a finite polymer chain will ultimately succeed in translocating through a pore.

In the more general framework of  anomalous diffusion, the presence of boundaries has been especially studied for non-Gaussian processes. For instance, L\'evy flights are Markovian superdiffusive processes whose increments obey a L\'evy stable (symmetric) law of index $0\le \mu\le 2$. The Hurst exponent is $H = 1/\mu$ \cite{noi}. By virtue of the Sparre Andersen  theorem \cite{sparre}, the persistence exponent is $\theta = 1/2$, independent of $\mu$. The Laplace Transform of the scaling function  $R_+(y)$ has been computed in \cite{zumofen} for a generic value of $\mu$. A scaling analysis of this Laplace Transform  shows that $R_+(y)$ 
behaves as $R_+(y)\sim y^{1/(2 \mu)}$ as $y\to 0$  (this in in agreement with the scaling relation $\phi=\theta/H$), while  far from the boundary the L\'evy-stable behavior is recovered.
 
An increasing interest is devoted to Gaussian processes with
self-affine anomalous displacements $\langle x^2(t)\rangle \sim t^{2
H}$ with $0<H<1$ \cite{kantor,krug,vannimenus,raoul,SliusarenkoEtAl}. Our current results apply only to fractional Brownian motion, i.e.\ self-affine Gaussian processes defined by the autocorrelation function (\ref{autocorr1}). In particular for fBm it is known that (i) the process has stationary increments, (ii) $\theta=1-H$, and (iii) $\phi=\theta/H$. For all  other Gaussian processes with Hurst exponent $H$, (i) the increments are non-stationary, (ii) $\theta \neq 1-H$ and we particularly emphasize that, (iii) no scaling relation is known between $\phi$ and $\theta$ (unlike in fBm where $\phi=\theta/H$).  Among such processes it is possible to show that the one, defined by the autocorrelation function
\begin{equation}
\langle x(t_1) x(t_2)\rangle \sim (t_1+t_2)^{2 H}- |t_1-t_2|^{2 H}
\end{equation}
 describes the subdiffusive behavoir of a tagged monomer in an elastic
 interface which initially was  flat \cite{krug}. For this process the
 persistence exponent is known only to first order in $\epsilon$
 \cite{krug}, whereas neither
 the exponents $\phi$, nor $\gamma $ are  known analytically. It would
 be interesting to determine the full scaling function $R_+(y)$ for
 this process within our perturbative framework.

\acknowledgements
It is a pleasure to thank Pierre Le Doussal, Mehran Kardar, and Andrea
Zoia for useful discussions. This work is supported by ANR grant 09-BLAN-0097-01/2.

\appendix
\section{The action} \label{a:action}

The aim of this Appendix is to determine the action ${\cal S}^{(1)}[x]$, the first correction to the Brownian action, ${\cal S}^{(0)}[x]$, in the expansion of ${\cal S}[x]$ in Eq.~(\ref{expans1}).
As a first step we expand the autocorrelation function (\ref{autocorr1}) around $H=1/2$, setting $D=1$,
\begin{eqnarray}\label{Kdef0}
\langle x(t_1) x(t_2) \rangle &=& G^{-1}(t_1,t_2)  \\
&=& [G^{(0)}]^{-1}(t_1,t_2)+ \epsilon \, K(t_1,t_2) +O(\epsilon^2).  \nn 
\end{eqnarray}
The first term is  the autocorrelation function for $H=1/2$,
\begin{equation}
\label{G01}
[G^{(0)}]^{-1}(t_1,t_2)= 2 \, \text{min}(t_1, t_2),
\end{equation}
the second term gives the correction at  first order in $\epsilon$,
\begin{equation}\label{A3}
 K(t_1,t_2)=2 \big[t_1 \log(t_1)+t_2 \log(t_2)-|t_1-t_2| \log |t_1-t_2| \big].
\end{equation}
Inverting Eq.~(\ref{Kdef0}) and expanding up to order $\epsilon$ one gets 
\begin{eqnarray}\label{a3}
G &=& G^{(0)}+\epsilon G^{(1)} +O(\epsilon^2)\\
G^{(1)}&=&-  G^{(0)} K G^{(0)}\ ,
\end{eqnarray}
where $G^{(0)}(t_1,t_2)$ is defined as
\begin{equation}
\label{condition}
\int_0^\infty \rmd t' G^{(0)}(t_1,t')[G^{(0)}]^{-1}(t',t_2)= \delta(t_1-t_2).
\end{equation}
One can check that  the kernel of the Brownian action, ${\cal S}^{(0)}[x]$, i. e.,
\begin{equation}
\label{G0}
G^{(0)} (t_1,t_2)= -\frac{1}{2}\, \delta'' (t_{1}-t_{2}), 
\end{equation}
satisfies Eq.~(\ref{condition}), namely,
\begin{align}
& -\frac{1}{2} \int_0^\infty \rmd t'\, \delta'' (t_{1}-t') [G^{(0)}]^{-1}(t',t_2) \nn \\
&=  -\int_0^\infty \rmd t'\, \delta'' (t_1-t') \text{min}(t',t_2) \nonumber \\
&= -\partial_{t_{1}}^{2} \min (t_{1},t_{2}) = \delta (t_{1}-t_{2})\ .
\end{align}
It remains to compute the term $G^{(1)}$. Integrating by parts one has
\begin{align}\label{f54}
&G^{(1)}(t_1,t_2)\nn \\
&=- \frac{1}{4} \int_0^t \int_0^t \rmd t'\, \rmd t''\, \delta'(t_1-t')  \delta'(t_2-t'')   \partial_{t'}\partial_{t''} K(t',t'')  \nonumber\\
&=\frac{1}{2} \int_0^t \int_0^t \rmd t'\, \rmd t''\, \delta'(t_1-t')  \delta'(t_2-t'')\nn \\
& \qquad \qquad \qquad \times  \partial_{t'} \partial_{t''}\left( |t'-t''| \log |t'-t''| \right)\ ,  
\end{align}
using that the first two terms in (\ref{A3}) do not contribute since
they only depend on one of the times. The derivative is
\begin{align}\label{a4}
&\partial_{t'} \partial_{t''}\left( |t'-t''| \log |t'-t''| \right)\nn \\
&\qquad  = - \frac{1}{|t'-t''|} -2 \delta(t'-t'')\big( 1+ \log|t'-t''|\big)\ .
\end{align}
The second term is not well-defined. We decide to introduce a
regularization for coinciding times $t=t' \rightarrow
\log|t-t'|=\log \tau$  where  $\tau>0$ should be thought of as the
time-discretization of the path-integral. Let us first give the final
result, before commenting on this approximation: 
\begin{eqnarray}\label{f54b}
\!\!\!\lefteqn{G^{(1)}(t_1,t_2)}\nn \\
&=&- \frac{1}{2} \int_0^t \int_0^t \rmd t'\, \rmd t'' \,
\delta'(t_1-t') \frac{1}{|t'-t''|}  \delta'(t_2-t'') \nn \\
&&  -2 (1+\log \tau) [G^{(0)}].  
\end{eqnarray} 
This yields for the action
\begin{eqnarray}\label{a5}
{\cal S}^{(1)}[x]&=&\int_0^t  d t_1 \int_0^t  d t_2 \, \frac12 x(t_1) G^{(1)}(t_1,t_2) x(t_2) \nn \\
&=& - \frac14 \int_0^t  d t_1 \int_0^t  d t_2 \, \frac{\partial_{t_1} x(t_1) \partial_{t_2} x(t_2)}{|t_1-t_2|} \nn \\
&& - 2\, {\cal S}^{(0)}[x] (1+\log \tau).
\end{eqnarray}
We see that the only possibly ambiguous term, the term  of order $\ln  \tau$, is proportional to
the zeroth-order action $ {\cal S}^{(0)}[x]$, thus equivalent to a
change in the diffusion constant $D$. Thus its effect is easy to check in the
final result, when looking at observables in a domain unaffected by
the boundary.

\section{Evaluation of $Z_+^{(1)}(x,t)$}
\label{s:Evaluation of Z+}
 
\subsection{Evaluation of $Z_+^A(x,t)$}
\label{s:Evaluation of Z+A}
This term is easily evaluated. Indeed, Eq.~(\ref{eqZA}) can be recast in the following form 
\begin{eqnarray}
\lefteqn{Z_+^A(x,t)}\nn\\
&=&-2(1+\log \tau) \nn \\
&& \times \lim_{x_0 \to 0}\frac{1}{x_0}  \left. \frac{\partial}{\partial a} \right|_{a=1}\int_{x(0)=x_0}^{x(t)=x} {\cal{D}}[x]  \rme^{- a {\cal S}^{(0)}[x]}\,  \Theta [x ]  \nonumber \\
&=& -2(1+\log \tau) \nn \\
&& \times \lim_{x_0 \to 0}\frac{1}{x_0}  \left. \frac{\partial}{\partial a} \right|_{a=1} \sqrt{\frac{a}{4 \pi t}}  \left[e^{-\frac{a}{4 t}(x-x_0)^2}-e^{-\frac{a}{4t}(x+x_0)^2}  \right] \nonumber \\
&=&  (1+\log \tau)\frac{x}{\sqrt{4 \pi t^3}} e^{-\frac{x^2}{4 t}}\left(\frac{x^2}{2 t}-3\right).
\end{eqnarray}
In going from the first to the second line we have used the expression
of the propagator in the Brownian case in Eq.~(\ref{Z+0}), introducing the
factor of $a$ from the observation that the latter appears together
with $x^{2}$, and readjusting the normalization.

In terms of the variable $z=x/\sqrt{2 t}$ this gives
\begin{equation}
\label{ZA}
Z_+^A(z,t)=   Z_+^{(0)}(z,t)A(z)
\end{equation}
where $Z_+^{(0)}(z,t)= z e^{-z^2/2}/ (\sqrt{2 \pi} t) $ is defined in (\ref{6}) and 
\begin{equation}
\label{Az}
A(z)=(1+\log \tau)\left(z^2 -3 \right)\ .
\end{equation}

\subsection{ $\tilde{Z}_+^B(x_0,x,s)$: The integration over $k$}
We split  $\tilde{Z}_+^B(x_0,x,s)$ into two parts
\begin{eqnarray}\label{km12}
\tilde{Z}_+^B(x_0,x,s)&=& \tilde{I}_1(x_0,x,s )+ \tilde{I}_2(x_0,x,s ) \\
{\tilde I}_1(x_0,x,s ) &=&     \frac4{x_0} \int\limits_{-\infty}^{\infty}\frac{\rmd
k}{2\pi} \left[\cos( k x\sqrt{s}  )- 
e^{-x \sqrt{s} }\right]\nn \\
&& \qquad \qquad\times  \left[\cos( k x_0 \sqrt{s}  )- 
e^{-x_0 \sqrt{s} }\right]\nn\\
&& \qquad \qquad \times\frac{ k^2 \log  (1+k^{2} )}{\sqrt{s} (1 + k^2)^2 }\label{km12b-2}
 \\
\tilde I_2(x_0,x,s ) &=&   \frac4{x_0} \int\limits_{-\infty}^{\infty}\frac{\rmd
k}{2\pi} \left[\cos( k x\sqrt{s}  )- 
e^{-x \sqrt{s} }\right]\nn \\
&& \qquad \qquad\times  \left[\cos( k x_0 \sqrt{s}  )- 
e^{-x_0 \sqrt{s} }\right]\nn\\
&& \qquad \qquad \times\frac{ k^2 \left[ \log (\tau s )+\gamma_{\mathrm E} \right]}{\sqrt{s} (1 + k^2)^2 }\label{km12c-2}
\end{eqnarray}

\subsubsection{$\tilde I_1(x_0,x,s )$}\label{a:tI1b}
The expansion of this term for small $x_0$ must be done with care;  when $x_0$ acts as a regulator, one cannot simply expand in it. 
We claim, and show below that 
\begin{equation}\label{B7}
\tilde I_1(x_0,x,s ) =   \tilde I_{1}^{A}(x_0,x,s ) +\tilde I_{1}^{B}(x,s ) +O(x_0)
\end{equation}
with
\begin{eqnarray}
\tilde I_{1}^{A}(x_0,x,s )&=& - \frac{4 e^{-x \sqrt{s} }}{x_0 \sqrt{s} }    \int\limits_{-\infty}^{\infty}\frac{\rmd
k}{2\pi}  \left[\cos( k x_0 \sqrt{s}  )- 1\right]\nn\\
&& \qquad \qquad \qquad \times \frac{ k^2 \log  (1+k^{2} )}{ (1 + k^2)^2 } \\
\tilde  I_{1}^{B}(x,s )&=& 4\int\limits_{-\infty}^{\infty}\frac{\rmd
k}{2\pi}  \left[  \cos( k x\sqrt{s}  ) -e^{-x \sqrt{s} }\right] \nn\\
&& \qquad \qquad \quad \times \frac{ k^2 \log  (1+k^{2} )}{ (1 + k^2)^2 }
\end{eqnarray}
In order to prove this, we group the four terms in (\ref{km12b-2}) into two times two terms;  the first combination is 
\begin{eqnarray}\label{B10}
&&\!\!\! -\frac{4\rme^{-x_0 \sqrt s}}{x_0}  \int\limits_{-\infty}^{\infty}\frac{\rmd
k}{2\pi} \left[\cos( k x\sqrt{s}  )- 
e^{-x \sqrt{s} }\right]\frac{ k^2 \log  (1+k^{2} )}{ (1 + k^2)^2 } \nn\\
&& =\left[-\frac{4}{x_0} + 4 \sqrt s + O(x_0)\right] \nn\\
&&\qquad \times \int\limits_{-\infty}^{\infty}\frac{\rmd
k}{2\pi} \left[\cos( k x\sqrt{s}  )- 
e^{-x \sqrt{s} }\right]  \frac{ k^2 \log  (1+k^{2} )}{ (1 + k^2)^2 } \nn \\
&& = \tilde I_1^{\mathrm{div}}(x_0,x,s)+  \tilde I_1^B(x,s) +O(x_0)
\end{eqnarray}
where the divergent contribution is 
\begin{eqnarray}
\tilde I_1^{\mathrm{div}}(x_0,x,s) &=& -\frac{4}{x_0}\int\limits_{-\infty}^{\infty}\frac{\rmd
k}{2\pi} \left[\cos( k x\sqrt{s}  )- 
e^{-x \sqrt{s} }\right] \nn\\
&& \qquad \qquad \times \frac{ k^2 \log  (1+k^{2} )}{ (1 + k^2)^2 }\ .
\end{eqnarray}
This expansion in $x_0$ is justified since $\frac{\rme^{-x_0 \sqrt s}}{x_0}$ stands outside the integrand, thus does not act as a regulator. 

The second contribution to (\ref{km12b-2}) is 
\begin{eqnarray}
&&\!\!\!\frac{4}{x_0}  \int\limits_{-\infty}^{\infty}\frac{\rmd
k}{2\pi} \left[\cos( k x\sqrt{s}  )- 
e^{-x \sqrt{s} }\right] \cos(k x_0\sqrt s)\frac{ k^2 \log  (1+k^{2} )}{ (1 + k^2)^2 } \nn\\
&&=\frac{2}{x_0} \int\limits_{-\infty}^{\infty}\frac{\rmd
k}{2\pi} \Big[\cos( k (x+x_0)\sqrt{s}  )+\cos( k (x-x_0)\sqrt{s}  )\nn\\
&&\qquad\qquad\qquad - 2
e^{-x \sqrt{s} }\cos(k x_0\sqrt s) \Big] \frac{ k^2 \log  (1+k^{2} )}{ (1 + k^2)^2 }
\end{eqnarray}
Since $x\gg x_0$, we can Taylor-expand $\cos( k (x+x_0)\sqrt{s}  )$ and $\cos( k (x-x_0)\sqrt{s}  )$, leading to 
\begin{eqnarray}
&&\frac{4}{x_0} \int\limits_{-\infty}^{\infty}\frac{\rmd
k}{2\pi} \Big[\cos( k x\sqrt{s}  )-
e^{-x \sqrt{s} }\cos(k x_0\sqrt s) \Big] \nn\\
&&\qquad \qquad\quad \times \frac{ k^2 \log  (1+k^{2} )}{ (1 + k^2)^2 } + O(x_0)\nn\\
&& = - 
\tilde I_1^{\mathrm{div}}(x_0,x,s) + 
\tilde I_1^{A}(x_0,x,s)+O(x_0)
\label{B13}
\end{eqnarray}
The contributions proportional to $\tilde I_1^{\mathrm {div}}$ cancel between (\ref{B10}) and (\ref{B13}), and we arrive at the decomposition (\ref{B7}).

We now treat the two contributions to (\ref{B7}). 
The first contribution $\tilde I_{1}^{A}(x_0,x,s )$ can be evaluated analytically. After integration over $k$ we find  a Bessel function, which can be
expanded in $x_0 $ as
\begin{align}\label{}
&\tilde I_{1}^{A}(x_0,x,s )= \\
&=  -4 e^{-\sqrt{s} x} \left[ \log (x_0 )+\frac{1}{2}\log (s)+\gamma_{\mathrm{E}} -1\right]+O (x_0) \nn
\end{align}

The second contribution $\tilde  I_{1}^{B}(x,s )$ can be evaluated using the relation
\begin{equation}
 \frac{k^2 }{(1+k^2)^2} \log(1+k^2)= \left[\frac{\rmd }{\rmd u}\bigg|_{u=1}- \frac{\rmd }{\rmd u}\bigg|_{u=0} \right]   \frac{1}{(1+k^2)^{u+1}}.
\end{equation}
We rewrite $ \tilde I_{1}^{A}(x,s )$ as
\begin{equation}
\tilde I_{1}^{B}(x,s )= 4  \left[\frac{\rmd }{\rmd u}\bigg|_{u=1}- \frac{\rmd
}{\rmd u}\bigg|_{u=0} \right]      \int\limits_{-\infty}^{\infty} \frac{\rmd
k}{2\pi}   \frac{e^{i k x \sqrt{s}} - e^{-x \sqrt{s} }}{(1+k^2)^{u+1}}
.
\end{equation}
It can be split into two parts,
\begin{align}
&\tilde I_{1}^{B}(x,s )=\tilde I_{1a}(x,s )+\tilde I_{1b}(x,s ) \\
&\tilde{I}_{1a}(x,s ) \nn \\
& ~~ =     -4\,  e^{-x \sqrt{s} }  \left[\frac{\rmd
}{\rmd u}\bigg|_{u=1}- \frac{\rmd }{\rmd u}\bigg|_{u=0} \right]     \int\limits_{-\infty}^{\infty} \frac{\rmd
k}{2\pi}   \frac{1}{(1+k^2)^{u+1}} \nn \\
& ~~= - \left[1+\log(4) \right] e^{-x \sqrt{s} }  \\ 
&\tilde{I}_{1b}(x,s ) \nn \\
& ~~=       \left[\frac{\rmd }{\rmd u}\bigg|_{u=1}- \frac{\rmd }{\rmd u}\bigg|_{u=0} \right]  4  \int\limits_{-\infty}^{\infty} \frac{\rmd
k}{2\pi}  e^{i k x \sqrt{s}}   (1+k^2)^{-(u+1)}\label{B11}
\end{align}
To do the $k$-integral in  $\tilde I_{1b} (x,s)$,
it is useful to introduce the integral representation
\begin{equation}\label{km12bb}
(1+k^2)^{-(u+1)} = \frac{1}{\Gamma(1+u)} \int_0^\infty  \rmd z\, z^u e^{-(1+k^2)z}. 
\end{equation}
This gives 
\begin{eqnarray}
\tilde{I}_{1b}(x,s ) &=& 4  \left[\frac{\rmd }{\rmd u}\bigg|_{u=1}- \frac{\rmd }{\rmd u}\bigg|_{u=0} \right]  \bigg[ \frac{1}{\Gamma(1+u)} \nn \\
&& \times \int_{-\infty}^{\infty}\frac{\rmd
k}{2\pi}   e^{i k x\sqrt{s}  } \int_0^\infty  \rmd z\, z^u e^{-(1+k^2)z} \bigg]\ ,\ \ \qquad 
\end{eqnarray}
and performing the Gaussian integral over $k$ yields
\begin{eqnarray}\label{km13}
\tilde{I}_{1b}(x,s ) &=& 4  \left[\frac{\rmd }{\rmd u}\bigg|_{u=1}- \frac{\rmd }{\rmd u}\bigg|_{u=0} \right]  \bigg[\frac{1}{\Gamma(1+u)} \nn \\
&&\qquad \times \int_0^\infty  \frac{\rmd z}{2 \sqrt{\pi}} z^{u-1/2} e^{-\frac{s x^2}{4 z} -z} \bigg]\ .\quad 
\end{eqnarray}

\subsubsection{$\tilde I_2(x_0,x,s )$}\label{a:tI2b}
$\tilde I_{2} (x_0,x,s)$ can be calculated using residue calculus. We use $x_0 <x$  to expand the expression,  choosing every pole in the
half-plane in which the corresponding exponential factor converges. The result is 
\begin{eqnarray}\label{}
\tilde I_{2}(x_0,x,s) =     \frac{\gamma_{\mathrm{E}} +\ln (\tau
s)}{2\sqrt{s} x_0 } &\bigg[ & \frac{ \sqrt{s} (x_0 +x)-1}{e^{\sqrt{s} (x+x_0)}} \ \ \qquad \nn  \\
&& \ -\frac{ \sqrt{s} (x-x_0 )-1}{e^{\sqrt{s} (x-x_0)}}
  \bigg]\qquad \  \
\end{eqnarray}
Expanding for small $x_0$ yields 
\begin{equation}\label{}
\tilde I_{2} (x_0,x,s) = e^{-\sqrt{s} x} \left(2-\sqrt{s} x\right) [\log (
\tau s )+\gamma_{\mathrm{E}} ]+O\left(x_0 \right)
\end{equation}

\subsubsection{Summary of all terms contributing to $\tilde Z_+^B(x_0,x,s)$ }\label{sub:SumI}
It is useful to re-organize 
\begin{eqnarray}
\tilde Z_+^B(x_0,x,s) &=& \tilde I_{1}^{A}(x_0,x,s ) + \tilde I_{1a}(x,s )+\tilde I_{1b}(x,s ) \nn\\&&+\tilde I_{2}(x,s ) + O(x_0)
\end{eqnarray}
 as the sum of three contributions:
\begin{equation}
\tilde Z_+^B(x_0,x,s)= \tilde J_0(x_0,x,s)+ \tilde J_1(x,s)+ \tilde J_2(x,s)+O(x_0)\ .
\end{equation}
The first term depends on $x_0$, 
\begin{equation}
\tilde J_0(x_0,x,s)= e^{-x \sqrt{s}} \left[ 3 -2 \gamma_{\mathrm{E}}  + 2 \log(  \tau /2 ) -4 \log (x_0)  \right]\ ,
\end{equation}
while the other two terms are
\begin{eqnarray}
\tilde J_1(x,s ) &=&  4  \left[\frac{\rmd }{\rmd u}\bigg|_{u=1}- \frac{\rmd }{\rmd u}\bigg|_{u=0} \right]  \bigg[\frac{1}{\Gamma(1+u)} \nn \\
&&\qquad \times \int_0^\infty  \frac{\rmd z}{2 \sqrt{\pi}} z^{u-1/2} e^{-\frac{s x^2}{4 z} -z} \bigg] \\
\tilde J_2(x,s)&=& - x \sqrt{s} e^{-x \sqrt{s}} \left[ \gamma_{\mathrm{E}} + \log( \tau s)    \right]\qquad \qquad 
\end{eqnarray}

\subsection{$Z_+^B(x,t)$: The inverse Laplace transform of $\tilde{Z}_+^B(x,s)$}
\label{AB2}

The inversion of $\tilde J_0(x_0,x,s)$ is done by  observing that
\begin{equation}
\tilde{Z}_+^{(0)} (x,s) = \lim_{x_0 \to 0}\frac{1}{x_0}  Z_+ (x_0,x,s) =e^{-\sqrt{s} x} \ .
\end{equation}
This yields 
\begin{equation}
\label{I1}
J_0(x_0,x,t)= Z_+^{(0)}(x,t) B_0(x_0)\ ,
\end{equation}
where $ Z_+^{(0)}(x,t)= \frac{x}{2 \sqrt{\pi} t^{3/2} } e^{-\frac{x^2}{4t}}$, and 
\begin{equation}
\label{B0}
 B_0(x_0)=3 -2 \gamma_{\mathrm{E}}  + 2 \log( \tau /2) -4 \log x_0\ .
\end{equation}

The inverse Laplace transform of the second term can be done directly, 
\begin{eqnarray}\label{a9}
J_{1}(x,t )&:=&  {\cal{L}}_s^{-1}\left[ \tilde{J}_{1}(x,s
)\right]  \nn \\
&&= 2 \left[\frac{\rmd }{\rmd u}\bigg|_{u=1}- \frac{\rmd
}{\rmd u}\bigg|_{u=0} \right] \bigg[ \frac{1}{\Gamma(1+u)} \nn \\
&&\qquad   \times  \int_0^\infty  \frac{\rmd z}{\sqrt{\pi}} z^{u-1/2} e^{-z} 
\delta\left(\frac{x^2}{4 z} -t\right)  \bigg]\ .\qquad 
\end{eqnarray}
We observe that $\delta\left(\frac{x^2}{4 z} -t\right)=\delta\left(\frac{x^2}{4 t} -z\right) z/t$ and obtain
\begin{equation}
J_1(x,t ) = \frac{2}{\sqrt{\pi} t} e^{-\frac{x^2}{4 t}}      \left[\frac{\rmd }{\rmd u}\bigg|_{u=1}- \frac{\rmd }{\rmd u}\bigg|_{u=0} \right]  \frac{\big(\frac{x^2}{4t}\big)^{u+1/2}}{\Gamma(1+u)}\ .
\end{equation}
Finally
\begin{eqnarray}\label{a10}
J_1(x,t ) &=&   \frac{1}{\sqrt{2 \pi} t} \frac{x}{\sqrt{2 t}} e^{-\frac{x^2}{4t}} 
     \nn \\
&&\times  \bigg[ \frac{x^2}{2 t} \left(\gamma_{\mathrm E}-1-\log 2 +\log \left(\frac{x^2}{2 t}\right) \right)\nn \\
&& \qquad \quad   -2 \left( \gamma_{\mathrm E} +\log\left(\frac{x^2}{2 t}\right)-\log 2 \right)   \bigg]\ .\qquad ~~
\end{eqnarray}
Introducing the variable  $z=x/\sqrt{2 t}$ we have:
\begin{equation}
\label{I1}
J_1(z,t )=     \frac{z}{\sqrt{2 \pi} t}  e^{-z^2/2} B_1(z)=Z_+^{(0)}(z,t) B_1(z)\ ,
\end{equation}
where $Z_+^{(0)}(z,t)= z \exp(-z^2/2)/\sqrt{2 \pi} t$, see Eq.~(\ref{6}),  and 
\begin{equation}
\label{B1}
B_1(z)= ( z^2-2)(\gamma_{\mathrm E} -1+2  \log z - \log 2)  -2 \ .
\end{equation}
This completes the Laplace inversion of 
$\tilde{J_1}(x,s )$.

\subsubsection{$\tilde J_2(x,s )$}\label{a:tI2}
The Laplace inversion of the second term $\tilde J_2(x,s )$ is more complicated, and we split it  as \begin{eqnarray}\label{km22}
\tilde J_2 (x,s) &=&- \big[\log (\tau s) +\gamma_{\mathrm E} \big] x \sqrt{s}  e^{-x \sqrt{s} }
 \nonumber \\
&=& x ( \log \tau  +\gamma_{\mathrm E} )  \frac{\rmd}{\rmd x}  e^{-x \sqrt{s} }+ x \frac{\rmd}{\rmd x} \left(\log s \, e^{-x \sqrt{s} }\right) \nonumber \\
&=&\tilde{J}_{2a} (x,s)+\tilde{J}_{2b} (x,s).
\end{eqnarray}
It is easy to perform the Laplace inversion of the first term:
\begin{eqnarray}\label{111}
 J_{2a} (x,t) &=&  (\log  \tau  +\gamma_{\mathrm E} )   \frac{x}{\sqrt{4 \pi t^3}} \frac{\rmd}{\rmd x} x e^{-\frac{x^2}{4 t}} \nonumber \\
 &=&   (\log  \tau  +\gamma_{\mathrm E} )  \frac{x}{\sqrt{4 \pi t^3}} e^{-\frac{x^2}{4 t}} \left( 1-\frac{x^2}{2 t}\right)\ .\qquad 
\end{eqnarray}
The inverse Laplace transform of the second term can be written as
\begin{equation}
J_{2b} (x,t)=x  \frac{\rmd}{\rmd x} f(x,t)\ ,
\label{uno}
\end{equation}
where
\begin{eqnarray}\label{due}
\int_0^\infty e^{-s t} f(x,t)  \rmd t &=& e^{-x \sqrt{s}} \log s=\tilde
g_{1} (s) \tilde g_{2} (s) \qquad \\
\tilde g_{1} (s)&=&  \sqrt{s} e^{-x \sqrt{s}} \\
\tilde g_{2} (s) &=&   \frac{\log s}{\sqrt{s}}\ .
\end{eqnarray}
The idea  is to inverse-Laplace transform $\tilde g_{1} (s)$ and
$\tilde g_{2} (s)$, and then to calculate $f (x,t)$ as convolution of
$g_{1} (t)$ and $g_{2} (t)$, using (\ref{conv-theorem}). These inverses are 
\begin{eqnarray}\label{a11}
 g_1(t) &=&  \frac{1}{2 \sqrt{\pi t^3}} \left( \frac{x^2}{2 t} -1 \right) e^{-\frac{x^2}{4 t}}  \\
 g_2(x,t)&=&   -\frac{\log(4 t)+\gamma_{\mathrm E}}{\sqrt{\pi t}}
\end{eqnarray}
The  convolution is 
\begin{eqnarray}\label{a12}
&& \!\!\!f(x,t)=\int_0^t g_1(t') g_2(t-t') \rmd{t'} \\
&&=-\int_0^t  \frac{\rmd{t'}}{2 \pi {t'}^{3/2} }\left(\frac{x^2}{2 t'}-1\right) e^{-\frac{x^2}{4 t'}} \,\frac{\log(4[t-t'])+\gamma_{\mathrm E}}{\sqrt{t-t'}}\ .\nn 
\end{eqnarray}
Using (\ref{uno}) we have
\begin{equation}
J_{2b} (x,t)=\frac{x^2}{4 \pi} \int_0^t  \frac{\rmd{t'}}{ {t'}^{5/2}} \left[\frac{x^2}{2 t'}-3\right] e^{-\frac{x^2}{4 t'}} \frac{\log(4[t-t'])+\gamma_{\mathrm E}}{\sqrt{t-t'}}
\label{finalea}
\end{equation}
Making a change of variables $t'=u t$, and using $z=x/\sqrt{2t}$, this gives 
\begin{eqnarray}
J_{2b} (z,t)&=&\frac{z^2}{2 \pi t} \int_0^1  \frac{\rmd{u}}{ {u}^{5/2} \sqrt{1-u}} \left[\frac{z^2}{ u}-3\right]\nn\\
&&\qquad \times e^{-\frac{z^2}{2 u}} \left[\log(4 t) +\gamma_{\mathrm E} +\log(1-u) \right]\qquad 
\label{finaleaa}
\end{eqnarray}
The integral contains two pieces, which we note
\begin{equation}
J_{2b} (z,t)=\frac{(\log(4 t)+\gamma_{\mathrm E}) F_2(z)+F_3(z)}{t} \ .
\end{equation}
The first piece is 
\begin{eqnarray}
 F_2(z)&:=&\frac{z^2}{2 \pi}\int_0^1\frac{\rmd{u}}{u^{5/2} \sqrt{1-u}} \left(\frac{ z^2}{u}-3\right) e^{-\frac{z^2}{2 u}}\nn\\
 & =&e^{-\frac{z^2}{2}}\frac{z}{\sqrt{ 2 \pi}}(z^2-1).
\end{eqnarray}
The second integral 
\begin{equation}
 F_3(z):=\frac{z^2}{2 \pi}\int_0^1\frac{\rmd{u}}{u^{5/2} \sqrt{1-u}} \log(1-u) \left(\frac{ z^2}{u}-3\right) e^{-\frac{z^2}{2 u}}
\end{equation}
is more difficult, but can be performed using Mathematica. A convenient substitution $\alpha=z^2(1/u-1)$ allows to write
\begin{equation}
\label{F3}
 F_3(z)= e^{-\frac{z^2}{2}} \frac{z}{\sqrt{2 \pi} } \,{\cal{I}}(z) \ ,
\end{equation}
where
\begin{eqnarray}\label{Iz}
    {\cal{I}}(z)&=& \frac{1}{\sqrt{2 \pi} z^2}\int_0^\infty
    \frac{\rmd{\alpha}}{\sqrt{\alpha}} \log\!\Big(\frac{\alpha}{z^2{+}\alpha}\Big)
    (z^2{+}\alpha) (z^2{+}\alpha{-}3) e^{-\frac \alpha 2} \nn \\
&=& \frac{z^{4}}{6} \, _2F_2\Big(1,1;\frac{5}{2},3;\frac{z^{2}}{2}\Big)+ \pi (1 -z^{2}) \text{erfi}(z/\sqrt{2})\nn \\
&& -3 z^{2}+ \sqrt{2\pi } e^{\frac{z^{2}}2} z+2\ .
\end{eqnarray}
$ \text{erfi}$ is the imaginary error-function, 
\begin{equation}\label{a14}
 \text{erfi}(x):=\frac{2}{\sqrt{\pi}}\int_{0}^{x}\rmd z\, \rme^{z^{2}}\ .
\end{equation} 
The hypergeometric function
$_2F_2\left(1,1;\frac{5}{2},3;{z^{2}}/{2}\right)$ can be defined by its series expansion
\begin{eqnarray}\label{a15}
_2F_2\Big(1,1;\frac{5}{2},3;\frac{z^{2}}{2}\Big) &=& 24
\sum_{n=0}^{\infty}  \frac{  n! (2 z^{2})^n}{ (2 n+4)!} \ .
\end{eqnarray}
The error-function and the exponential function can be combined in another
 converging series, 
\begin{eqnarray}\label{}
\lefteqn{e^{\frac{z^2}{2}} z-\sqrt{\frac{\pi }{2}} \left(z^2-1\right)
   \text{erfi}\Big(\frac{z}{\sqrt{2}}\Big)} \nn \\
&&=- \sum_{n=0}^{\infty}  \frac{2^{1-n} z^{2 n+1}}{(2 n-1) (2 n+1) n!}
\end{eqnarray}
While problems of numerical precision appear for $y>7$, we can use the
asymptotic expansion 
\begin{equation}\label{Iasymp}
{\cal I} (z) = 1-\gamma_{\mathrm{E}}-\log (2  z^{2}) +\frac{1}{2
   z^2}-\frac{1}{2 z^4}+\frac{5}{4
   z^6}+O ( z^{-8} )
\end{equation}
At $z=7$, the relative numerical agreement of
(\ref{Iasymp}) and (\ref{Iz}) is about $10^{-6}$. 

Note that $\int_{0}^{\infty}\rmd z\, z \rme^{-z^{2}/2} {\cal I}
(z)=0$, thus ${\cal I} (z)$ does not contribute to the normalization. 

\subsubsection{$J_{2}(x,t)= J_{2a}(x,t )+J_{2b}(x,t )$}\label{}
The sum $J_{2}(x,t)= J_{2a}(x,t )+J_{2b}(x,t )$ can  be expressed 
using the variable $z=x/\sqrt{2 t}$ as
\begin{equation}
\label{I2}
J_2(z,t )=    Z_+^{(0)}(z,t) B_2(z,t)
\end{equation}
where $Z_+^{(0)}(z,t)=z e^{-z^2/2}/(\sqrt{2 \pi}\, t) $ and 
\begin{equation}
\label{B2}
B_2(z)=\left(z^2-1\right)\log( 4t/\tau) +{\cal{I}}(z).
\end{equation}

\subsection{Summary of all terms}\label{sub:Sum}
In summary, 
\begin{eqnarray}
\label{Z1}
Z_+^{(1)}(z,t)=  Z_+^{(0)}(z,t) &\!\Big[\!&A(z) +B_0(x_0)+B_{1} (z)+B_{2} (z) \nn\\
&& - a_1 \log x_0 \Big]
\end{eqnarray}
where $Z_+^{(0)}(z,t)= z e^{-z^2/2}/ (\sqrt{2 \pi} t) $ is defined in Eq.\
(\ref{6}).
The terms in question are given in Eqs.\ (\ref{Az}), (\ref{B0}), (\ref{B1}) and
(\ref{B2}), and repeated here:
\begin{eqnarray}\label{AzRep}
A(z)&=&(1+\log \tau)\left(z^2 -3 \right) \\ 
\label{B0Rep}
B_0(x_0)&=&3 -2 \gamma_{\mathrm{E}}  + 2 \log(\tau /2) -4 \log x_0\\
\label{B1Rep}
B_1(z) 
&=& ( z^2-2)(\gamma_{\mathrm E} -1+2  \log z - \log 2)  -2 \qquad 
\\
\label{B2Rep}
B_2(z)&=&\left(z^2-1\right)\log( 4t/\tau) +{\cal{I}}(z).
\end{eqnarray}
Their sum is 
\begin{eqnarray}\label{A+B1+B2}
\lefteqn{A(z) +B_0+B_{1}(z)+B_{2} (z) - a_1 \log x_0}\nn \\
&=&\left\{ ( z^2-2) \left[  \log (2 z^{2 }t) +\gamma_{\mathrm E}
\right]-2 \right\} +{\cal I} (z)\nn\\
&&  -(4+a_1)\log x_0+c (t) \nn  \\
c (t)&=& \log ( t)+2-2 \gamma_{\mathrm{E}}\  .\qquad 
\end{eqnarray}
The result is arranged such that the term in the curly brackets, when multiplied by $Z_{+}^{(0)} (z,t)$,
integrates to zero, as does $Z_{+}^{(0)} (z,t)\, {\cal I} (z)$. The propagator $Z_+^{(1)}(z,t)$ becomes independent of $x_0$ if $a_1=-4$, equivalent to  $\phi_0=1-4\epsilon+O(\epsilon^2)$. As expected, $\phi_0=\phi$, see Eq.~(\ref{final3}).

Since 
$c(t)$ only contributes to the (time-dependent) normalization, it does not enter the scaling function $R_+(y)$. 

On the other hand, the only contribution  to the normalization of the propagator $Z_+(x,t)$ comes from $c(t)$. Since 
$Z_{+}^{(0)} (x,t)$ integrated over $x$ equals 1, we
conclude that the survival-probability is 
\begin{eqnarray}\label{}
S (x_0,t) &=& t^{-\frac{1}{2}} \left[1+ \epsilon \Big(2 -
2\gamma_{\mathrm{E}} +\ln t \Big) \right]\nn \\
& \sim& t^{-\theta}\ , \qquad
\theta =\frac{1}{2}- \epsilon + O(\epsilon^2)
\end{eqnarray}
in agreement with $\theta =1-H$. This is a non-trivial check of our calculations.

\section{Scaling arguments}
\label{s:scaling}
Consider a process $x (t')$, starting at $x (0)= x_{0}$, and
arriving at $x$ at time $t$, 
without having crossed zero, i.e.\ $x (t')>0$ for all
$t'<t$. Denote 
$Z_{+} (x_{0},x,t) $ its arrival probability density at $x$. Further denote 
\begin{equation}\label{g1}
S(x_{0},t) := \int_{0}^{\infty}\rmd x\, Z_{+} (x_{0},x,t) 
\end{equation}
the survival probability or the persistence up to time $t$. At late times and fixed $x_0$,
for many processes, this survival probability decays algebraically 
\begin{equation}\label{g2}
S (x_{0},t) \sim  t^{-\theta}\ ,
\end{equation}
where $\theta$ is the persistence exponent \cite{current}. Let us now assume that the 
process $x(t)$ is self-affine. This simply means that the process is characterized 
by a single growing length scale $x\sim t^{H}$ where $H$ is the Hurst exponent
of the process. For example, ordinary Brownian motion is a self-affne process
with $H=1/2$. Since the only length scale is $x\sim t^H$,  the survival
probability $S(x_0,t)$ is a function of only the scaled variable $y_0=x_0/t^H$, i.e,
$S(x_0,t)= G\left(\frac{x_0}{t^H}\right)$. In order that $S(x_0,t)\sim t^{-\theta}$ for large $t$ and fixed $x_0$, 
the scaling function $G(y)$, for small $y$, must behave as
\begin{equation}\label{g4}
G(y_0)\sim y_0^{\phi} \ , \qquad {\rm where}\,\,\, \phi=\frac{\theta}{H}\ .
\end{equation}

We next define $p_{x_0}(x,t)$ as the conditional probability density of finding the walker, given that it has not been absorbed at any previous time:
\begin{equation}
p_{x_0}(x,t)=\frac{Z_+(x_{0},x,t)}{\int_{0}^{\infty} \rmd x\, Z_+(x_{0},x,t)} = \frac{Z_+(x_{0},x,t)}{S(x_0,t)}\ .
\label{conditional}
\end{equation}
Note that following Eq.~(\ref{4}), the probability distribution of a
non-adsorbed particle   is for $x_0 \to 0$
\begin{equation}
P_+(x,t) =p_{0}(x,t)=  \lim_{x_0 \to 0}\frac{Z_+ (x_0,x,t)}{\int_0^\infty \rmd x \,Z_+ (x_0,x,t)}.
\end{equation}
We anticipate the following scaling form for $Z_+(x_0,x,t)$
\begin{equation}\label{scaling_1}
Z_+(x_0,x,t)= \frac{1}{t^{H}} F \left(\frac{x_{0}}{t^{H}}, \frac{x}{t^{H}}\right).
\end{equation}
In terms of the scale variables $y=x/t^H$ and $y_0=x_0/t^H$ we get from (\ref{conditional}) and (\ref{scaling_1})
\begin{equation}
F(y,y_0)=G(y_0)p_{y_0}(y)
\label{FY}
\end{equation}
where $p_{y_0}(y)$ is the conditional probability density
(\ref{conditional}) expressed in terms of the rescaled variables. In
the long-time limit, $y_0 \to 0$ and $F(y,y_0)$ can be factorized as  
\begin{equation}
F(y,y_0) \sim y_0^{\theta/H} p_{0}(y)= y_0^{\theta/H} R_{+}(y) .
\label{FY2}
\end{equation}
Let us now consider the limit $y \to 0$ and suppose that $p_{0}(y)=
R_{+}(y) \sim y^\phi$. The process is time-reversible invariant, since its
increments are stationary, i.e., a path from $x_0$ to $x$ forward in
time plays the same role as a path from $x$ to $x_0$ backward in
time. As a consequence, $F (y,y_{0})$ is a symmetric function,  $F
(y,y_{0})= F (y_{0},y)$. Factorization of probabilities for $x$ and
$x_{0}$ to zero and symmetry thus implies  $F(y,y_0) \sim (y_0 y)^{\theta/H}$ and it follows the proposed scaling relation $\phi=\theta/H$.


\begin{thebibliography}{0}

\bibitem{persistence} B. Derrida, A.J. Bray, and C. Godr\`eche, J. Phys. A 27, 
L357 (1994); A.J. Bray, B. Derrida, and C. Godr\`eche, Europhys. Lett. 
27, 175 (1994); B. Derrida, V. Hakim, and V. Pasquier, Phys. Rev. Lett. 
75, 751 (1995).
 
\bibitem{redner} S. Redner, {\em A Guide to First-Passage Processes}, Cambridge University Press, New York, (2001).
\bibitem{current} S.N. Majumdar, {\em Persistence in Nonequilibrium Systems}, Curr. Sci. {\bf 77}, 370 (1999).
\bibitem{marcos} M. Marcos-Martin, D. Beysens, J.-P. Bouchaud, C. 
Godr\`eche, and I. Yekutieli, Physica D {\bf 214}, 396 (1995). 
\bibitem{yurke} B. Yurke, A.N. Pargellis, S.N. Majumdar, and C. Sire, 
Phys. Rev. E  {\bf 56}, R40 (1997). 
\bibitem{tam} W.Y. Tam, R. Zeitak, K.Y. Szeto, and J. Stavans, Phys. 
Rev. Lett. {\bf 78}, 1588 (1997). 
\bibitem{walsworth} G. P. Wong, R. W. Mair, R. L. Walsworth, and D. G. Cory Phys. Rev. Lett. {\bf 86}, 4156 (2001).
\bibitem{dougherty} D. B. Dougherty et al., Phys. Rev. Lett. {\bf 89}, 136102 (2002).
\bibitem{leeuwen} J.MJ. Leeuwen, V.W.A. de Villeneuve, H.N.W. Lekkerkerker J. Stat. Mech.: Theory and Experiment, P09003 (2009).
\bibitem{stavans}  J. Soriano, I. Braslavsky, D. Xu, O. Krichevsky, and J. Stavans, Phys. Rev. Lett.   {\bf 103}, 226101 (2009).
\bibitem{diffusionm} S.N. Majumdar, C. Sire, A.J. Bray, and S.J. Cornell, 
Phys. Rev. Lett. {\bf 77}, 2867 (1996); B. Derrida, V. Hakim, 
and R. Zeitak, ibid. 2871.

\bibitem{satyapert} S. N. Majumdar, C. Sire, Phys. Rev. Lett. {\bf 77}, 1420 (1996).

\bibitem{braypert} K. Oerding, S. J. Cornell, A. J. Bray, Phys. Rev. E {\bf 56}, R25 (1997).

\bibitem{global} S.N. Majumdar, A.J. Bray, S.J. Cornell, and C. Sire, 
Phys. Rev. Lett. {\bf 77}, 3704 (1996);  S.N. Ma jumdar and A. J. Bray, Phys. Rev Lett.  {\bf 91}, 030602 (2003).


\bibitem{hilhorst} H. Hilholrst, Physica A {\bf 277}, 124(2000).


\bibitem{krug} J. Krug, H. Kallabis, S. N. Majumdar, S. J. Cornell, A. J. Bray, and C. Sire, Phys. Rev. E {\bf 56}, 2702 (1997).

\bibitem{cornell} S.N. Majumdar, and S.J. Cornell, Phys. Rev. E {\bf 57}, 3757 
(1998). 
\bibitem{brayinterface}  S.N. Majumdar and A. J. Bray, Phys. Rev Lett. {\bf 86}, 3700 (2001).

\bibitem{ref1} J. J. Kasianowicz, E. Brandin, D. Branton, and D. W. Deamer, Proc. Natl. Acad. Sci. U.S.A. {\bf 93}, 13770 (1996).

\bibitem{ref2} A. Meller, L. Nivon, and D. Branton, Phys. Rev. Lett. {\bf 86}, 3435 (2001).

\bibitem{ref3} A. Meller, J. Phys: Cond. Matter. {\bf 15}, R581 (2003).

\bibitem{ref4} A. Yu. Grosberg, S. Nechaev, M. Tamm, and O. Vasilyev, Phys. Rev. Lett. {\bf 96}, 228105 (2006).

\bibitem{ref5} M. Muthukumar, J. Chem. Phys. {\bf 111}, 10371 (1999).

\bibitem{ref6} D. K. Lubensky and D. R. Nelson, Biophys. J. {\bf 77}, 1824 (1999).

\bibitem{ref7} A. J. Storm, C. Storm, J. Chen, H. Zandbergen, J. F. Joanny, and C. Dekker, Nano Lett. {\bf 5}, 1193 (2005).

\bibitem{ref7bis} D. Panja and G.T. Barkema, J. Chem. Phys. {\bf 132}, 014902 (2010); D. Panja and G.T. Barkema, R.C. Ball, J. Phys.: Condens. Matter {\bf 19} (2007) 432202.

 
\bibitem{kardar1} Y. Kantor and M. Kardar, Phys. Rev. E {\bf 76}, 061121 (2007).


\bibitem{kardar2} C. Chatelain, Y. Kantor and M. Kardar, Phys. Rev. E {\bf 78}, 021129 (2008).

\bibitem{phi1}  A. Zoia, A. Rosso, S. N. Majumdar, Phys. Rev. Lett. {\bf 102}, 120602 (2009). 


\bibitem{pgd} P.-G. de Gennes, {\em Scaling Concepts in Polymer Physics} (Cornell Univ. Press, Ithaca, NY, 1979).


\bibitem{LeDoussalWiese2008a}
P.~Le Doussal and K.J. Wiese,
\newblock Phys. Rev. E {\bf 79} (2009)   051105.


\bibitem{LeDoussalMiddletonWiese2008}
P.~{Le~Doussal}, A.A. Middleton  and K.J.\ Wiese,
\newblock Phys. Rev. E  {\bf 79} (2009)   050101 (R).

\bibitem{LeDoussalWiese2008c}
P.~{Le~Doussal} and K.J. Wiese,
\newblock Phys. Rev. E {\bf 79} (2009)   051106.


\bibitem{LeDoussalWiese2009a}
P.~{Le Doussal} and K.J. Wiese,
\newblock Phys. Rev. E {\bf 82} (2010)   011108.


\bibitem{LeDoussalWieseMoulinetRolley2009}
P.~Le Doussal, K.J. Wiese, S.~Moulinet  and E.~Rolley,
\newblock EPL {\bf 87} (2009)   56001.


\bibitem{LeDoussalWiese2008b}
P.~{Le Doussal} and K.J. Wiese,
\newblock EPL {\bf 86} (2009)   22001.

\bibitem{PLD1}
P.~{Le Doussal}, J. Stat. Mech. (2009) P07032.

\bibitem{PLD2}
G.~Schehr, and 
P.~{Le Doussal}, J. Stat. Mech. (2010) P01009.

\bibitem{simulation} T. Dieker, http://www2.isye.gatech.edu/$\sim$adieker3/fBm/ 

\bibitem{Garcia}
R. Garcia-Garcia, A. Rosso, G. Schehr, Phys. Rev. E {\bf 81}, 010102(R) (2010).



\bibitem{hitting} S. N. Majumdar, A. Rosso, A. Zoia, Phys. Rev. Lett. {\bf 104}, 020602 (2010). 




\bibitem{noi} A.~Zoia, A.~Rosso, and M.~Kardar, Phys. Rev. E {\bf 76} 021116 (2007).

\bibitem{sparre} E.~Sparre Andersen, Math. Scand. {\bf 1}, 263 (1953); Math. Scand. {\bf 2}, 195 (1954).

\bibitem{zumofen} G.~Zumofen and J.~Klafter, Phys. Rev. E {\bf 51}, 2805 (1995).
\bibitem{vannimenus}
A. Rosso, W. Krauth, P. Le Doussal, J. Vannimenus, and K.J. Wiese,
Phys. Rev. E {\bf 68} (2003) 036128. 


\bibitem{raoul}
R. Santachiara, A. Rosso and W. Krauth, J. Stat. Mech. (2007) P02009.

\bibitem{kantor}  A. Amitai, Y.  Kantor, M.  Kardar, Phys. Rev. E {\bf 81}, 011107  (2010). 

\bibitem{Barkai}
L. Lizana, T. Ambjornsson, A. Taloni, E. Barkai, M.A. Lomholt,
Phys. Rev. E {\bf 81}, 051118 (2010). 

\bibitem{SliusarenkoEtAl}
O.Y. Sliusarenko, V.Y. Gonchar, A.V. Chechkin,
I.M. Sokolov, and R. Metzler, Phys. Rev. E {\bf 81}, 041119 (2010).


\end{thebibliography}
\end{document}